\documentclass{article}
\begin{document}
\begin{center}
{\huge{ {\bf Thermodynamic Partition Function of Matrix 
Superstrings}
}}\\
\vskip 0.2truein
{\bf G. Grignani}\footnote{Permanent address: 
{\it Dipartimento di Fisica and Sezione
I.N.F.N., Universit\`a di Perugia, Via A. Pascoli I-06123 Perugia,
Italia}. E-mail gianluca.grignani@PG.infn.it   Research supported by I.N.F.N. 
and M.U.R.S.T. of Italy.}
and {\bf G. W. Semenoff}\footnote{Permanent address: {\it Department of Physics
and Astronomy, University of British Columbia, 6224 Agricultural Road, 
Vancouver, British Columbia, Canada V6T 1Z1}. E-mail: semenoff@alf.nbi.dk  
Research supported by NSERC of Canada and the Niels Bohr Fund of Denmark.}
\vskip 0.15truein
{\it The Niels Bohr Institute\\
Blegdamsvej 17\\
DK-2100 Copenhagen 0\\
Denmark}

\vskip 0.5truein
{\bf Abstract}
\vskip 0.1truein
\end{center}

We show that, in the limit of zero string coupling, $g_s \rightarrow 0$,  
the thermodynamic partition function of matrix string
theory is identical to that of the finite temperature, discrete
light-cone quantised (DLCQ) type IIA superstring.    
We discuss how the superstring is recovered in 
the decompactified $R^+\rightarrow\infty$ limit.

 \def\be{\begin{equation}}
\def\ee{\end{equation}}
\def\bea{\begin{eqnarray}}
\def\eea{\end{eqnarray}}
\def\nn{\nonumber}
\def\const{{\rm const}}
\def\v{\varphi}
\def\s{\sigma}
\def\vcl{\varphi_{\rm cl}}
\newcommand{\no}[1]{:\!#1\!:}
\def\la{\left\langle}
\def\ra{\right\rangle}
\def\d{\partial}
\def\se{S_{\rm eff}}
\def\tr{\rm Tr}
\section{Introduction and summary}

M-theory is an eleven dimensional theory whose compactification on a circle
is equivalent to ten dimensional IIA superstring theory with 
coupling constant $g_s$ related to the radius of the compact dimension.
The higher Kaluza-Klein modes corresponding to M-theory degrees of freedom 
with non-zero momenta in the compact direction are the D0-branes 
of the IIA theory.  
According to the proposal of Banks, Fischler, Shenker and Susskind \cite{bfss} 
the full dynamics of M-theory in the infinite momentum frame 
is described by the large N limit of supersymmetric matrix quantum
mechanics with gauge group U(N). There is a further conjecture that, at finite
$N$, the matrix model describes discrete light-cone quantised (DLCQ) 
M-theory, the de-compactified limit occurring when $N$ and the light-cone
compactification radius are taken to infinity in the appropriate 
way \cite{su}.  

Compactification of the matrix model on a spatial circle,
to yield a description of IIA superstring 
theory by a two-dimensional super-Yang-Mills theory, was discussed by several
authors \cite{setsus,m,tbns,dvv}.
Dijkgraaf, Verlinde and Verlinde \cite{dvv} gave arguments to show how  
IIA superstrings emerge in the limit of weak string coupling.
In their scenario, strings are associated with the diagonal components 
of the $N\times N$ matrices which decouple from the off-diagonal components
in the $g_s\rightarrow 0$ limit.  The infrared dynamics of the diagonal 
elements are described by 
an ${\cal N}=8$ supersymmetric sigma model with orbifold target space 
$\left( R^8\right)^N/S_N$.
On the other hand, the non-interacting 
first-quantised superstring is described in the light-
cone gauge by the supersymmetric sigma model with target space $R^8$.  
The idea of ref. \cite{dvv} is that, when $N$ is large, the quantum states of
the conformal 
field theory with target space $\left( R^8\right)^N/{S_N}$ are identical 
to those of 
the second quantised superstring theory which lie in the appropriately
symmetrised infinite direct
product of Hilbert spaces of conformal field theories each with 
target space $R^8$.  This idea is supported by the identification of 
elliptic genera on the orbifold $({\cal M})^N/S_N$ for compact
Kahler manifold ${\cal M}$ and similar objects which are expected to 
arise from second quantisation of a superstring theory on 
${\cal M}\times S^1$, which 
was conjectured in \cite{dvv1} and proven in \cite{dmvv}.

In this Paper we shall present an alternative way of seeing how the matrix
model produces the quantum states of the string 
theory.  We shall compare the thermodynamic partition function of the 
matrix string theory in the limit $g_s\rightarrow 0$ with the thermodynamic 
partition function of the non-interacting 
discrete light-cone quantised (DLCQ) 
IIA superstring. We shall find 
that, at least in this non-interacting limit, the two are identical. 
This implies that, at zero string 
coupling (which means infinite Yang-Mills coupling), 
the spectrum of the matrix model
is identical to the spectrum of all multi-string states of the
non-interacting  
discrete-light-cone quantised IIA superstring theory.  This identity 
requires neither taking the large $N$ limit nor the de-compactification of
the light-cone, $R^+\rightarrow\infty$.  
  
The natural extension of our result to the interacting theory would be
to show that the quantum states of the second quantised interacting 
DLCQ IIA string theory with total longitudinal momentum $N/R^+$ 
would be identical to the gauge invariant eigenstates of the Hamiltonian of 
$U(N)$ matrix string theory.  This is the content of the conjecture
in \cite{su}.  
It should be possible to check this idea by comparison of perturbative  
corrections to the thermodynamic partition functions of each model.  
Superstring perturbation theory has been discussed in the context of matrix
string theory by several authors 
\cite{wyn1,wyn2,ghv,bon1,bon2,bon3}.  It is likely that some of 
these proposals could be studied systematically within our approach.

The partition function of matrix string theory should be constructed 
by using the Hamiltonian of the appropriately compactified matrix
theory, 
\bea
{\cal H}=\frac{R^+}{2}\int_0^1d\sigma_1\tr\left[\Pi_i^2+\frac{1}{4\pi^2
\alpha'^2}(D_{1} X^i)^2 -\frac{1}{4\pi^2 \alpha'^3 g_s^2}
\left[X^i,X^j\right]^2
+\frac{E^2}{g_s^2\alpha'} \right.\cr\left. -  
\frac{1}{2\pi \alpha'^{3/2} g_s}\psi^T\gamma_i\left[X^i,\psi\right]-
\frac{i }{2\pi\alpha'}\psi^T \gamma\cdot D_{1}\psi\right]
\label{mtheoryham}
\eea
where the notation and conventions are explained in Section 3.
This Hamiltonian is to be interpreted as the light-cone Hamiltonian of 
M-theory
\begin{equation}
{\cal H}=P^+
\label{pplus}
\end{equation}
where M-theory has  
two compactified dimensions, one of them spatial, in order that it describes 
the IIA superstring and the other the  
light-cone direction $X^+$ with radius $R^+$.  
The gauge invariant states of this 
Hamiltonian are those of M-theory in a sector of fixed longitudinal
momentum
\begin{equation}
P^-=\frac{N}{R^+}
\label{pminus}
\end{equation}
which is discrete since momentum in a compact direction should be quantised
in units of inverse radius. To obtain the theory on un-compactified 
ten dimensional space the limit $R^+\rightarrow\infty$ would  
eventually have
to be taken.\footnote{Note that in the matrix model description of M-theory, 
N is the number of D0-branes which are the basic building blocks of all 
objects.  In the matrix string theory on the other hand, because
of the 9-11 flip which is used in (\cite{dvv}), N labels the quantum of
momentum $P^+$.}

We use (\ref{pplus}) and (\ref{pminus}) to 
form the thermodynamic partition function,
\begin{eqnarray}
Z[ \beta]={\rm Tr}\left( e^{-\beta P^0}\right)
={\rm Tr}\left( e^{-\frac{\beta}{\sqrt{2}}\left( P^++P^-\right)}\right)
\label{pf1}\\
=\sum_{N=0}^\infty
e^{-\beta N/\sqrt{2}R^+}{\rm Tr}_N\left(e^{-\beta{\cal H}/\sqrt{2}}
\right)
\label{pf22}
\end{eqnarray}
where in (\ref{pf1}) 
the trace is to be taken over all eigenstates of both $P^+$ and $P^-$ 
and $\beta=1/k_B T$ with $k_B$ the Boltzmann constant and $T$ the temperature.
The trace over $P^-$ leads to the summation over $N$ in (\ref{pf22}).
Taking the trace over gauge invariant 
states of the D=2 super-Yang-Mills theory with fixed $N$
in (\ref{pf22}) yields an expression for
the partition function in terms of a Euclidean functional integral
\begin{equation}
Z[\beta]=\sum_{N=0}^\infty 
\int [dA][dX^i][d\psi]~\exp\left(-\beta N/\sqrt{2}R^+-S_E[A,X^i,
\psi]\right)
\label{pf2}
\end{equation}
where the action $S_E$ is the 
continuation to two-dimensional Euclidean space of the 
action from which the Hamiltonian (\ref{mtheoryham}) can be obtained
by canonical quantisation.  The two dimensional integration in 
$S_E[A,X^i,\psi]$ is over a torus with 
spatial coordinate $\sigma_1\in[0,1)$ and temporal coordinate in $\sigma_2
\in[0,\beta/\sqrt{2})$ and the fields have periodic boundary 
conditions 
in the spatial directions, $X^i$ and $A_\alpha$ are periodic and
$\psi$ are anti-periodic in Euclidean time.  By suitable rescaling, 
the Euclidean time interval can be set to $[0,1)$ and the temperature 
would then
appear in coefficients in the action. Details are discussed in Section 3.
Note that, as is usually the case, the
thermal boundary conditions, which are different for Fermi and Bose degrees
of freedom, break the supersymmetry of the theory.  This 
reflects the different thermal populations of fermionic and bosonic states 
of the matrix model (as well as the IIA superstring) and is an expected
feature of a supersymmetric theory at finite temperature.  For a recent 
discussion of supersymmetry and other issues in 
matrix theory at finite temperature, see \cite{ams,mak1,mp}.

We then consider this partition function in the limit $g_s\rightarrow 0$.  
In that limit, it reduces to a model where the degrees of freedom are the
diagonal elements of the matrix-valued fields.  The gauge fields decouple
and the scalar and fermion fields form a super-conformal field theory with
target space the orbifold $(R^8)^N/S_N$.   
We shall find that, for this model, the thermodynamic free energy, defined by 
$$
F[\beta]=-\frac{1}{\beta}\ln Z[\beta]
$$
is given by
\begin{eqnarray}
F&=&-\frac{1}{\beta}
\sum_{N=0}^{\infty}
\sum_{\stackrel{r|N}{r\ odd}}\sum_{s=0}^{N/r-1}~
\frac{1}{N}\exp\left(-\frac{N\beta}{\sqrt{2}R^+}\right)
\cdot\cr &\cdot &
\int d\vec X d\psi^i \exp\left[-
\frac{1}{4\pi\alpha'}\int\sqrt{g}\left(g^{\alpha\beta}
\partial_\alpha \vec X\cdot \partial_\beta \vec X-i2\pi\alpha'\bar\psi
\gamma^ae_a^\alpha\partial_\alpha\psi \right)\right]
\label{matstpf}
\end{eqnarray}
where the odd 
integers $r$ are the odd divisors of N.  The super-conformal
field theory defined by the path integral has the light-cone gauge fixed 
Green-Schwarz superstring action. The
worldsheet is a Euclidean torus with coordinates $0\leq\sigma^1,\sigma^2<1$
and metric
\begin{equation}
g_{\alpha\beta}=\left( \matrix{   1 & \tau_1\cr \tau_1& 
\vert\tau\vert^2\cr}\right)~=~e_\alpha^a \delta_{ab}e_\beta^b
\label{metric}
\end{equation}
In the path integral in (\ref{matstpf}), the Bose fields $\vec X$ have periodic
boundary conditions in both the worldsheet space and time directions
whereas the Fermi fields $\psi$ have periodic and anti-periodic 
boundary conditions in the space and time directions, respectively.
$\tau=\tau_1+i\tau_2$ is the
usual Teichm\"uller parameter of the torus.  In (\ref{matstpf}) it takes
the discrete values
\begin{equation}
\tau=\frac{s}{N/r}+ i \frac{\beta R^+}{\sqrt{2}2\pi\alpha'}\frac{r}{N/r}
\label{tau}
\end{equation}

We shall find that the IIA superstring partition function, using
discrete light-cone quantisation, and 
where the compactification radius of $X^+$ is
also  $R^+$, is given by an expression {\it identical} to (\ref{matstpf}).
In DLCQ of the IIA superstring, 
the Teichm\"uller parameter $\tau_1$, which is summed over the
values
$$\tau_1=
0,\frac{1}{N/r},\frac{2}{N/r},\ldots,\frac{N/r-1}{
N/r}
$$
originates as a Lagrange multiplier which enforces the Virasoro constraint
\begin{equation}
\sum_{s=0}^{N/r-1}\exp\left( 2\pi i\frac{s}{N/r}\left(
h-\tilde h\right)\right)
=\left\{ \matrix{ 1&~~{\rm if~}h-\tilde h=0~{\rm mod}~N/r\cr 0
&~~{\rm otherwise} }
\right.
\end{equation}
which, we shall argue in Section 2, is the appropriate constraint for 
the DLCQ superstring.
The other Teichm\"uller parameter $\tau_2$ arises from the summation over
eigenvalues of $P^+$.  On the other hand, 
in the matrix string theory, these parameters originate
from the sum over permutations which occur in the boundary conditions
of the diagonal components of the matrices when they are the 
variables of the orbifold sigma model.  Similar permutations have been argued 
to play a key role in the computation of the Witten index 
\cite{shat1,kos1,kos2,kostov} and in 
Yang-Mills theory partition functions on tori \cite{dadda,dadda1}.   
They are also important in obtaining the spectrum of two dimensional
Yang-Mills theory from the path integral \cite{dadda2,gpss} and in the 
correspondence between two dimensional Yang-Mills theory and a random 
surface model \cite{grotay1,grotay2,grotay3}.  

Since the thermodynamic partition function encodes information about
the energy levels and degeneracies of states, we conclude that the 
spectrum of the $g_s\rightarrow 0$ limit of 
matrix string theory is identical to the spectrum of the second-quantised 
non-interacting DLCQ IIA superstring.

It is interesting to study how 
the expression (\ref{matstpf}) will recover the superstring 
partition function in the limit $R^+\rightarrow\infty$. The mechanism
by which this occurs is clear.  The only suppression of the magnitude of $N$ 
in the summation in (\ref{matstpf}) comes from the exponential which has 
$R^+$ in the denominator.  When $R^+>>\beta$, the sum is dominated 
by very large $N$.  When $N$ becomes very large, $\tau$ in (\ref{tau})
should become a continuous variable.  
$\tau$ should correctly be defined as varying as the 
integers $s$ and $N/r$ vary. Then, when $N$ is very large, 
the increments in $\tau$ as we go between 
successive values of $s$ and $N/r$
are small and the summation over $s$, $N$ and $r$ become integrations
over $\tau_1$ and $\tau_2$ and a summation over $r$.
In fact, $\tau_1$ is varied by changing $s$ while holding $r$ and $N/r$ 
fixed, so that 
$$d\tau_1= \frac{(s+1)}{N/r}-\frac{s}{N/r}=\frac{1}{N/r}
$$
and $\tau_2$ is varied by changing $N/r$ but keeping $r$ and $s$ 
fixed\footnote{It is clear that this means changing $N$ to the next 
larger integer which is divisible by $r$},
$$
\frac{d\tau_2}{\tau_2^2}=-
d\left( \frac{1}{\tau_2}\right)=
\frac{\sqrt{2}2\pi\alpha'}{\beta R^+}\frac{1}{r}
\left(
\left(\frac{N}{r}+1\right)-\left(\frac{N}{r}\right)\right)
$$ 
$$
=\frac{\sqrt{2}2\pi\alpha'}{\beta R^+}\frac{1}{r}
$$
so that
$$
\frac{1}{N}=
\beta\frac{2\pi R^+}{4\pi^2\alpha'\sqrt{2}}
\frac{d\tau_1 d\tau_2}{\tau_2^2}
$$
With this identification, the continuum limit of the summations in
(\ref{matstpf}) is
\begin{equation}
\sum_{N=0}^{\infty}\frac{e^{-\frac{N\beta}{2R^+}}}{N}
\sum_{\stackrel{r|N}{r\ odd}}\sum_{s=0}^{N/r-1}\longrightarrow
\frac{\beta R^+}{\sqrt{2}2\pi\alpha'}
\int_0^1 d\tau_1\int_0^\infty 
\frac{d\tau_2}{\tau_2^2}\sum^\infty_{\stackrel{r=1}{r\ odd}}
e^{-\frac{r^2\beta^2}{4\pi\alpha'\tau_2}}
\end{equation}
and the result is identical to the expression for the thermodynamic free
energy of the superstring
\begin{eqnarray}
F&=&-
\frac{\sqrt{2}\pi R^+}{4\pi^2\alpha'}
\int_0^1 d\tau_1\int_0^\infty 
\frac{d\tau_2}{\tau_2^2}\sum^\infty_{\stackrel{r=1}{\rm r\ odd}}
e^{-\frac{r^2\beta^2}{4\pi\alpha'\tau_2}}\cdot
\cr &\cdot&
\int d\vec X d\psi^i \exp\left[-\frac{1}{4\pi\alpha'}
\int\sqrt{g}\left(g^{\alpha\beta}
\partial_\alpha \vec X\cdot \partial_\beta \vec X-i 2\pi\alpha'\bar\psi
\gamma^ae_a^\alpha\partial_\alpha\psi \right)\right]
\end{eqnarray}
Integrating the remaining variables and taking into account
that the space volume is given by $V=V_T\cdot \sqrt{2}\pi R^+$, 
with $V_T$ and $\sqrt{2}\pi R^+$ the volume of the transverse 
space and the length of the longitudinal direction, produces
the well-known expression for the free energy density~\cite{aw,a,ao,b}
\bea
\frac{F}{V}&=&-\int^{1/2}_{-1/2}\frac{2^7 d\tau_1}{(4\pi^2\alpha')^{5}}
\int_0^{\infty}
\frac{d \tau_2}{\tau_2^{6}}\left|\theta_4\left(0,2\tau\right)
\right|^{-16}\cdot\cr&\cdot&
\left[\theta_3\left(0,\frac{i\beta^2}{4\pi^2\alpha'\tau_2}\right)-
\theta_4\left(0,\frac{i\beta^2}{4\pi^2\alpha'\tau_2}\right)\right]
\eea

In Section 2 we will review the derivation of the finite temperature partition
function of the superstring.  We use the relativistic particle and the bosonic
string to illustrate the main idea in a simpler context and to derive some
necessary formulae.  Then we derive an 
expression for the free energy of the type IIA superstring using discrete 
light-cone quantisation, as well as the partition function of the superstring.

In Section 3 we discuss the thermodynamic partition function for the matrix 
string model. We argue that taking the $g_s\rightarrow 0$ limit reduces
the model to one for the diagonal components of the matrices with boundary
conditions that are (anti-)periodic up to permutations of the eigenvalues
in both the space and compact Euclidean time directions.  We then show how
the combinatorics of permutations reduces to a sum over coverings of
the torus, which produces a sum over the Teichm\"uller parameters 
in (\ref{matstpf}).

\subsection{Notation}

In this section, we summarise some of the notation. 
For a D-vector $x^\mu$, we shall use the convention for the Minkowski metric
\begin{equation}
ds^2=dx^\mu g_{\mu\nu} dx^\nu\equiv (dx^0)^2-{d\vec x}^{~2}
=2dx^+ dx^--\left( d\vec x_T\right)^2
\end{equation}
where $\vec x$ is the vector made from the $D-1$ spatial components of
$x^\mu$, the light-cone coordinates are
\begin{equation}
x^+=\frac{1}{\sqrt{2}}\left( x^0+x^{D-1}\right)
~~,~~~
x^-=\frac{1}{\sqrt{2}}\left( x^0-x^{D-1}\right)
\end{equation}
and  
$
\vec x_T=\left(x^1,\ldots,x^{D-2}\right)
$.

The Jacobi theta functions are
\begin{eqnarray}
\theta_1(\nu,\tau)=i\sum_{n=-\infty}^\infty (-1)^nq^{(n-1/2)^2/2}z^{n-1/2}
~,~
\theta_3(\nu,\tau)=\sum_{n=-\infty}^\infty q^{n^2/2}z^{n}
\\
\theta_2(\nu,\tau)=\sum_{n=-\infty}^\infty q^{(n-1/2)^2/2}z^{n-1/2}
~,~
\theta_4(\nu,\tau)=\sum_{n=-\infty}^\infty (-1)^nq^{n^2/2}z^{n}
\end{eqnarray}
where 
$$
q=\exp\left(2\pi i\tau\right)~~~,~~~z=\exp\left(2\pi i\nu\right)
$$
We shall also denote by $$\tilde
\theta_k(\nu,\tau)=\sum_{\stackrel{n=-\infty}
{n\neq0}}^\infty \ldots$$ 
the theta function where the $n=0$ term is absent from the sum.
The modular transformation properties are
\begin{eqnarray}
\theta_1(\nu,\tau+1)=e^{i\pi/4}\theta_1(\nu,\tau)~~&,&~~
\theta_2(\nu,\tau+1)=e^{i\pi/4}\theta_2(\nu,\tau)\cr
\theta_3(\nu,\tau+1)=\theta_4(\nu,\tau)~~&,&~~
\theta_4(\nu,\tau+1)=\theta_3(\nu,\tau)
\label{modular1}
\end{eqnarray}
and
\begin{eqnarray}
\theta_1(\nu/\tau, -1/\tau)=-\left(-i\tau\right)^{1/2}e^{\pi i\nu^2/\tau}
\theta_1(\nu,\tau) \cr
\theta_2(\nu/\tau, -1/\tau)=-\left(-i\tau\right)^{1/2}e^{\pi i\nu^2/\tau}
\theta_4(\nu,\tau) \cr
\theta_3(\nu/\tau, -1/\tau)=-\left(-i\tau\right)^{1/2}e^{\pi i\nu^2/\tau}
\theta_3(\nu,\tau) \cr
\theta_4(\nu/\tau, -1/\tau)=-\left(-i\tau\right)^{1/2}e^{\pi i\nu^2/\tau}
\theta_2(\nu,\tau) 
\label{modular2}
\end{eqnarray}
They obey Jacobi's abstruse identity
\begin{equation}
\theta_3^4(0,\tau)-\theta_4^4(0,\tau)-\theta_2^4(0,\tau)=0
\label{abstruse}
\end{equation}
and Jacobi's triple product formulae are
\begin{eqnarray}
\theta_1(\nu,\tau)=2q^{1/8}\sin(\pi\nu)\prod_{n=1}^\infty
\left(1-q^n\right)\left(1-2q^n\cos(2\pi\nu)+q^{2n}\right) \cr
\theta_2(\nu,\tau)=2q^{1/8}\cos(\pi\nu)\prod_{n=1}^\infty
\left(1-q^n\right)\left(1+2q^n\cos(2\pi\nu)+q^{2n}\right) \cr
\theta_3(\nu,\tau)=\prod_{n=1}^\infty
\left(1-q^n\right)\left(1-2q^{n-1/2}\cos(2\pi\nu)+q^{2n-1}\right) \cr
\theta_4(\nu,\tau)=\prod_{n=1}^\infty
\left(1-q^n\right)\left(1-2q^{n-1/2}\cos(2\pi\nu)+q^{2n-1}\right) \cr
\label{triple}
\end{eqnarray}

The Dedekind eta function is
\begin{equation}
\eta(\tau)=q^{1/24}\prod_{n=1}^\infty\left( 1-q^n\right)
\end{equation}
and has the modular transformation properties
\begin{equation}
\eta(\tau+1)=e^{i\pi/12}\eta(\tau)~~,~~\eta(-1/\tau)=(-i\tau)^{1/2}
\eta(\tau)
\end{equation}

\section{Thermodynamic free energy of the discrete light-cone 
quantised superstring}

 In this section, we will derive the thermodynamic free energy of the DLCQ 
IIA superstring.  To illustrate the point, we begin with a review of the 
relativistic particle and the bosonic string.  Then we discuss the DLCQ 
bosonic string at finite temperature.  Finally, we derive the partition 
function of the DLCQ superstring.

\subsection{Relativistic particle}

To begin, we 
consider the free energy per unit volume of a relativistic particle,
in $D$ spacetime dimensions
\begin{equation}
\frac{F}{V}= (-1)^f\frac{1}{\beta}\int \frac{ d^{D-1}p}{(2\pi)^{D-1}} 
\ln\left( 1-(-1)^f e^{-\beta\omega(\vec p)}\right)
\label{freenergy}
\end{equation}
$f=0$ if the particle is a boson and $f=1$ for a fermion.
The energy is
$\omega(\vec p)=
\sqrt{{\vec p}^2+M^2}
$ with $M$ the mass.

To write the integral (\ref{freenergy}) in terms of light-cone momenta
we use the identity
\begin{eqnarray}
\delta(p^0-\omega(\vec p))&=&2p^0\theta(p^0)\delta(p^2-M^2)
 \cr &=& \sqrt{2}\left(
 p^++p^-\right)\theta(p^++p^-)\delta(2p^+p^--{\vec p_T}^{~2}
 -M^2)\label{delta}
\end{eqnarray}
to write
\bea
\frac{F}{V}&=& \frac{(-1)^f}{\beta}\int \frac{dp^+dp^- d^{D-2}p_T}{\sqrt{2}\pi
(2\pi)^{D-2}}
\left( p^++p^-\right)\theta(p^++p^-)
\delta(2p^+p^--{\vec p_T}^{~2}
-M^2)\cr &\cdot &\ln\left( 1-(-1)^f
e^{-\frac{\beta}{\sqrt{2}}(p^++p^-)}\right)
\label{freenergy1}
\eea
which, upon integration over $p^+$, becomes
\begin{equation}
\frac{F}{V}= (-1)^f\frac{1}{\beta}\int_0^\infty \frac{dp^-}{\sqrt{2}\pi}\int 
\frac{d^{D-2}p_T}{(2\pi)^{D-2}}
\ln\left( 1-(-1)^f e^{-\frac{\beta}{\sqrt{2}}
\left(\frac{{\vec p_T}^{~2}+M^2}{2p^-}+p^-\right)}\right)
\label{freenergy2}
\end{equation}
Here, in order to equate the two terms which arise from
(\ref{freenergy1}), we have made use of a symmetry of the integrand
under an interchange of $p^+$ and $p^-$.

We then use a Taylor expansion of the logarithm to present the free
energy per unit volume as
\be
\frac{F}{V}= -\sum_{n=1}^\infty\frac{(-1)^{(n+1) f}}{n\beta}\int_0^\infty 
\frac{dp^-}{\sqrt{2}\pi}
~e^{-n\beta p^-/\sqrt{2}}\int \frac{d^{D-2}p_T}{(2\pi)^{D-2}}~
e^{-\frac{n\beta}{\sqrt{2}}\left(\frac{{\vec p_T}^{~2}+M^2}{2p^-}
\right)}
\label{freenergy3}
\ee

In this form, the free
energy is a trace over the space of states of a density matrix which
is made from the light-cone
Hamiltonian,
\begin{equation}
H_{\rm l.c.}=\frac{ {\vec p_T}^{~2}+M^2}{2p^-}
\label{lch}
\end{equation}
as
\begin{equation}
{\rm Tr}\left( e^{-\frac{n\beta}{\sqrt{2}}H_{\rm l.c.}}
\right)=V_T\int \frac{d^{D-2}p_T}{(2\pi)^{D-2}}~
e^{-\frac{n\beta}{\sqrt{2}}\left(\frac{{\vec p_T}^{~2}+M^2}{2p^-}
\right)}
\label{propagator}
\end{equation}
and which can be thought of as the trace of the light-cone 
time evolution operator over intervals of Euclidean time
$
n\beta/\sqrt{2}
$.  The volume is $V=LV_T$ with $V_T$ and $L$ the volume of the
transverse space and length of the longitudinal direction, respectively.

The Gaussian integral over $\vec p_T$  in (\ref{freenergy3}) 
can be done explicitly to get
\be
\frac{F}{V}=-\sum_{n=1}^\infty \int_0^\infty \frac{dp^-}{p^-} 
(-1)^{(n+1)f}\left( \frac{p^-}{\sqrt{2}\pi n\beta}\right)^{D/2}
\exp\left[-\frac{n\beta}{\sqrt{2}}\left(\frac{M^2}{2p^-}+
p^-\right)\right]
\label{lcpf}
\ee

Finally, let us take note of the modification of the formula
(\ref{lcpf}) which would be necessary in DLCQ.  The light-cone
direction $X^+$ is assumed to be compactified, so that we should 
identify $X^+$ and $X^++2\pi R^+$.  This identification gives the 
momentum conjugate to $X^+$, i.e. $P^-$ a discrete spectrum.
Thus, rather than the integral over $p^-$ in (\ref{lcpf}) we
should have a sum over the spectrum 
$$
p^-=k/R^+
$$
\be
\frac{F}{V}=-\sum_{n=1}^\infty \sum_{k=1}^\infty \frac{(-1)^{(n+1)f}}{k} 
\left(\frac{k}{\sqrt{2}\pi n\beta R^+}\right)^{D/2}
\exp\left[-\frac{n\beta}{\sqrt{2}}\left(\frac{R^+ M^2}{2k}+
\frac{k}{R^+}\right)\right]
\label{dlcpf}
\ee

This formula will be used to derive the path integral  
expression for  the free energy 
of the DLCQ closed string. An expression similar to it was quoted in
ref \cite{mop}.  

\subsection{DLCQ Bosonic closed string}

The equations of motion and constraints of the Bosonic closed string theory
are
\begin{eqnarray}
\partial_+\partial_- X^\mu =0 \\
\partial_+X^\mu\partial_+ X_\mu=0 \\
\partial_-X^\mu\partial_- X_\mu=0
\end{eqnarray}
$X^+$ is identified with $X^++2\pi R^+$. 
In closed string theory, boundary conditions are periodic,
\begin{equation}
X^\mu(\tau,\sigma+2\pi)=X^\mu(\tau,\sigma)+\delta^\mu_+2\pi R^+r
\end{equation}
where $r$ is an integer which is the number of times the string world-sheet
wraps the compact direction $X^+$.  

The equations of motion have the light-cone gauge solution
\begin{eqnarray}
X^-(\tau,\sigma)&=&x^-+P^-\tau   \cr
X^+(\tau,\sigma)&=&x^++P^+\tau + rR^+\sigma + ({\rm oscillators})  \cr
\vec X_T(\tau,\sigma)&=&\vec x_T + \vec P_T\tau + ({\rm oscillators}) 
\end{eqnarray}
Since the canonical commutation relations are
$$
\left[ x^+,P^-\right]=-i
$$
and $x^+$ is a compact variable, the momentum $P^-$ is quantised as
$$
P^-= \frac{k}{R^+}
$$
In this light-cone gauge, the Virasoro constraints can be solved to
eliminate the light-cone components of the oscillators, 
$\alpha^+_{n\neq 0}$ and $\tilde\alpha^+_{n\neq 0}$.  The remaining 
constraint which must be imposed is 
$$
L_0=0=\tilde L_0
$$
The sum of these gives the mass-shell condition,
\begin{equation}
0=L_0+\tilde L_0~
\rightarrow 
P^\mu P_\mu=M^2=\frac{2}{\alpha'}\left( h+\tilde h-2\right)
\label{spectrum}
\end{equation}
and their difference gives the condition
\begin{equation}
0=L_0-\tilde L_0~\rightarrow~ P^-\cdot(rR^+)=kr=h-\tilde h
\label{const}
\end{equation}
where, the Hamiltonians are
\begin{equation}
h=\sum_1^\infty {\vec\alpha}_{-m}\cdot{\vec\alpha}_m
\end{equation}
and 
\begin{equation}
\tilde h=\sum_1^\infty {\vec{\tilde\alpha}}_{-m}\cdot {\vec{
\tilde\alpha}}_m
\end{equation} 
The right- and left-moving 
light-cone gauge oscillators  have the algebra 
\begin{equation}
\left[ {\alpha}^i_m, {\alpha}^j_n\right]
=m\delta_{m+n}\delta^{ij}
\end{equation}
\begin{equation}
\left[ {\tilde \alpha}^i_m, {\tilde \alpha}^j_n\right]
=m\delta_{m+n}\delta^{ij}
\end{equation}
where $i,j=1,...,24$. (For the bosonic string, we fix the 
spacetime dimension to be D=26.) 

Now, we can examine the free energy of the bosonic string. At the one loop 
level, it is gotten by computing the sum of free energies of the particles 
in the string spectrum.  This is 
obtained from (\ref{dlcpf}) with $f=0$, 
using the fact that for the string $M^2$ are the eigenvalues of 
the operator given in 
(\ref{spectrum}) which also obey the constraint (\ref{const}).  
The constraint can be enforced by a discrete sum,
\begin{equation}
\sum_{s=0}^{k-1}\frac{1}{k}\exp\left(2\pi i\frac{s}{k}(h-\tilde h)\right)=
\left\{ \matrix{ 1 & {\rm ~if~}h-\tilde h=0~{\rm mod}~k\cr 0&~
{\rm otherwise} }\right.
\end{equation}
The trace 
over the string spectrum constrained by (\ref{const}) can be written as
\begin{eqnarray}
\frac{F}{V}=-\sum_{n=1}^\infty \sum_{k=0}^\infty \frac{1}{k} 
\left( \frac{k}{\sqrt{2}\pi n\beta R^+}\right)^{13}~
\sum_{s=0}^{k-1}\frac{1}{k}
\tr~
\exp\left(-\frac{n\beta R^+}{\sqrt{2}}
\frac{M^2}{2k}-\frac{nk\beta}{\sqrt{2}R^+}
\right. \cr \left.
+2\pi i 
\frac{s}{k}(\tilde{h}-h)\right)
\label{42}
\end{eqnarray}
We introduce the notation
\begin{equation}
\tau(n,k,s)=\tau_1+i\tau_2~,~~\tau_1(k,s)=\frac{s}{k}~,~~
\tau_2(k,n)= \frac{n\beta R^+}
{2\pi\alpha'\cdot\sqrt{2}k}
\label{taudefn}
\end{equation}
Later, we shall see that these parameters become the Teichm\"uller parameters
of the worldsheet of the string which is a torus.  We see that, for the DLCQ 
string, they take discrete values, which should merge to form a continuum of
tori when the limit $R^+\rightarrow\infty$ 
is taken.  In terms of these parameters, the
free energy of the Bosonic string is
\begin{eqnarray}
\frac{F}{V}=-\sum_{n=1}^\infty \sum_{k=1}^\infty \sum_{s=0}^{k-1}
\frac{1}{k^2} \left(\frac{1}{4\pi^2\alpha'\tau_2}\right)^{13}
e^{4\pi\tau_2-n^2\beta^2/4\pi\alpha'\tau_2}
\tr\left(e^{-2\pi i\bar\tau h}~e^{2\pi i \tau \tilde h}\right)
\label{fb}
\end{eqnarray}
The trace over the string states can be computed by noting that
$\vec{\alpha}_{m}$ are related to conventionally normalised harmonic 
oscillator operators
by 
\be
\vec{\alpha}_{m}=\sqrt{m}\, {\vec {a}}_m\ ,\quad\quad
\vec{\alpha}_{-m}=\sqrt{m}\, {\vec {a}^\dag}_m
\ee
so that
\be
\tr \left(x^{-\sum_{m=1}^{\infty} m\vec{a}_m^\dag\cdot\vec{a}_m}\right)=
\left(\prod^\infty_{m=1}\frac{1}{1-x^m}\right)^{24}
\ee
Then the free energy density reads
\begin{eqnarray}
\frac{F}{V}=-\sum_{n=1}^\infty \sum_{k=1}^\infty \sum_{s=0}^{k-1}
\frac{1}{k^2} \left(\frac{1}{4\pi^2\alpha'\tau_2}\right)^{13}~
\exp\left(-\frac{n^2\beta^2}{4\pi\alpha'\tau_2}\right)
\left| \eta\left(\tau\right)\right|^{-48}
\label{fb1}
\end{eqnarray}
This is the free energy density of the DLCQ bosonic string.  It is similar 
to the known free energy except that, rather than an integration over the 
Teichm\"uller parameters of the torus with the appropriate gauge fixed measure
of the Polyakov path integral, there is a discrete summation with values of
$\tau(n,k,s)$ given in (\ref{taudefn}).

Of course, it should recover the well known expression \cite{pol} for the
partition function of the Bosonic closed string when the light-cone is
de-compactified, i.e. in the limit $R^+\rightarrow\infty$.  One can see 
from equation (\ref{42}) that,
in this limit, the sum is dominated by combinations of integers where 
$nk$ is very large.  The correct continuum limit is gotten by considering
integration over all values of 
$\tau_2$ as the limit of doing the sum over the integers $k$ 
while holding $n$ and $s$ fixed.  
Then 
$$
-\tau_2 d\left(\frac{1}{\tau_2}\right)=
\tau_2(k,n)\left( \frac{1}{\tau_2(k,n)}
-\frac{1}{\tau_2(k+1,n)}\right)=\frac{1}{k}
$$
Furthermore, the integration over $\tau_1$ arises from the sum over $s$, 
holding $k$ and $r$ fixed, 
$$
d\tau_1(k,s)\equiv\tau_1(k,s+1)-\tau_1(k,s)
= \frac{s+1}{k}-\frac{s}{k}=\frac{1}{k}
$$
so that
$$
\frac{d^2\tau}{\tau_2}=\frac{1}{k^2}
$$
accounts for the factor of $1/k^2$ in (\ref{fb1}).
In the limit where the summations over $k$ and $s$ become continuous
integrals over $\tau_1$ and $\tau_2$ 
according to the formula
\begin{equation}
\sum_{k=0}^\infty \sum_{s=0}^{k-1}\frac{1}{k^2}
~\longrightarrow ~
\int_{-1/2}^{1/2}d\tau_1 \int_0^\infty \frac{1}{\tau_2}d\tau_2
\end{equation}
and $n$ is summed independently of
these variables, the usual expression for the free energy density of the
bosonic string is recovered,
\be
\frac{F}{V}=-\frac{1}{2(4\pi^2\alpha')^{13}}\int^{1/2}_{-1/2}
d\tau_1\int_0^{\infty}
\frac{d \tau_2}{\tau_2^{14}}\left|\eta(\tau)\right|^{-48}
\tilde\theta_3\left(0,i\beta^2/4\pi^2\alpha'\tau_2\right)
\ee
Note that this coincides with the result given by Polchinksi \cite{pol}
and agrees with the one quoted by Atick and Witten
\cite{aw} (see also  \cite{a,ao,b}) when one
uses the SL(2,Z) modular group to introduce a second integer in the 
thermal summation and restricts the integration in $\tau$ over the
fundamental domain of the torus
\be
-\frac{1}{2}<\tau_1<\frac{1}{2}\ ,\quad \tau_2>0\ ,\quad |\tau|>1\ .
\ee 

We further note that the free energies of both the DLCQ and de-compactified
bosonic strings can be presented in a form of a path integral over
coordinates transverse to the light-cone, 
\begin{eqnarray}
\frac{F}{L}&=&-\sum_{n=1}^\infty \int^{1/2}_{-1/2}\frac{d\tau_1}{4\pi^2\alpha'}
\int_0^\infty \frac{d\tau_2}{\tau_2^2}~
e^{-\frac{\beta^2 n^2}{4\pi\alpha'\tau_2}} \cr &\cdot&
\int d\vec X \exp\left( -\frac{1}{4\pi\alpha'}\int d^2\sigma\sqrt{g}
g^{\alpha\beta}
\partial_\alpha \vec X\cdot \partial_\beta \vec X\right)
\label{pif}
\end{eqnarray}
The functional integral defines a conformal field theory
on the transverse space $R^{24}$ with worldsheet  the 
torus with $0\leq\sigma^1,\sigma^2<1$ and metric (\ref{metric}).

\subsection{DLCQ IIA superstring}

Let us now consider the case of a closed type II 
superstring in ten dimensions.
The superstring theory contains equal numbers of space-time bosons and 
space-time fermions which are treated differently in the formula 
(\ref{dlcpf}).  We can get the superstring partition function by using the
average of the two contributions $f=0$ and $f=1$ in (\ref{dlcpf}) and
then summing over the total multiplicities of all allowed states of the
superstring theory, weighted by the appropriate Boltzmann factor.  This 
technique was originally used in \cite{a}.  The free energy is
\begin{equation}
\frac{F}{V}=-\sum_{\stackrel {n=1} {n ~\rm odd} }^\infty
\sum_{k=1}^\infty
\frac{2}{k}\left( \frac{1}{4\pi^2\alpha'\tau_2}\right)^5
e^{-n^2\beta^2/4\pi\alpha'\tau_2}\sum_Me^{-\pi\alpha'\tau_2 M^2}
\label{lcpfs}
\end{equation}
with $\tau_2$ defined in (\ref{taudefn}).
The mass spectrum for a closed superstring is given by the spectrum
of the operator
\be
M^2=\frac{2}{\alpha'}\sum_{i=1}^8 \sum_{n=1}^\infty n_i\left(h_{n_i}^B+
h_{n_i}^F+\tilde{h}_{n_i}^B+\tilde{h}_{n_i}^F\right)
\label{suspec}
\ee
where $h_{n_i}^B$, $h_{n_i}^F$, $\tilde{h}_{n_i}^B$ and $\tilde{h}_{n_i}^F$
are the number operators for each species of left- and right-moving worldsheet
boson and fermion degree of freedom.
The longitudinal momentum is given by 
\be
P^-=\frac{k}{R^+}
\ee
As in the case of the DLCQ bosonic string, the constraint $L_0-\tilde L_0
\sim 0$ implies that the oscillators numbers obey 
\be
\sum_{i=1}^8 \sum_{n=1}^\infty n_i \left(h_{n_i}^B+
h_{n_i}^F-\tilde{h}_{n_i}^B-\tilde{h}_{n_i}^F\right)=0~{\rm mod}~k
\label{suscons}
\ee
The free energy for the superstring is obtained by 
introducing in (\ref{lcpfs}) the trace over the string spectrum 
(\ref{suspec}) and implementing the constraint (\ref{suscons}) 
with a discrete Fourier sum. We use the formula for the trace over
number operators 
\begin{eqnarray}
\tr ~\left[e^{-2\pi\tau_2(h^B+h^F+\tilde h^B+\tilde h^F) + 
2\pi i\tau_1 (h^B+h^F-\tilde h^B-\tilde h^F)}\right]=
2^8\prod_{n=1}^{\infty}\left|
\frac{1+e^{2\pi i \tau n}}{1-e^{2\pi i \tau n}}\right|^{16}\cr
\equiv 2^8 \left|\theta_4\left(0,2\tau\right)\right|^{-16}
\label{strace}
\end{eqnarray}
where $\tau$ is given in (\ref{taudefn}) 
and the factor $2^8$ is due to the 
degeneracy of the fermionic ground states. 

Inserting this result into (\ref{lcpfs}), the free energy for the 
closed superstring then reads
\begin{equation}
\frac{F}{V}=-\sum_{\stackrel {n=1} {n ~\rm odd} }^\infty
\sum_{k=1}^\infty
\sum_{s=0}^{k-1}\frac{1}{k^2}\left( \frac{1}{4\pi^2\alpha'\tau_2}\right)^5
2^9 \left|\theta_4\left(0,2\tau\right)\right|^{-16}
e^{-n^2\beta^2/4\pi\alpha'\tau_2}
\label{lcpfs1}
\end{equation}
This is the free energy of the DCLQ type II superstring.

In the limit $R^+\rightarrow\infty$, again using the assumption that the 
summation over Teichm\"uller parameters goes to an integral over those 
parameters according to the formula
\begin{equation}
\sum_{k=0}^\infty \sum_{s=0}^{k-1}\frac{1}{k^2}
~\longrightarrow ~
\int_{-1/2}^{1/2}d\tau_1 \int_0^\infty \frac{1}{\tau_2}d\tau_2
\end{equation}
this expression reduces to the
free energy of the type II superstring,
\bea
\frac{F}{V}=-\frac{2^7}{\left(4\pi^2\alpha'\right)^5}
\int^{1/2}_{-1/2}d\tau_1\int_0^{\infty}
\frac{d \tau_2}{\tau_2^{6}}~~\left|\theta_4\left(0,2\tau\right)
\right|^{-16}\cdot\cr\cdot
\left[\theta_3\left(0,i\beta^2/4\pi^2\alpha'\tau_2\right)-
\theta_4\left(0,i\beta^2/4\pi^2\alpha'\tau_2\right)\right]
\label{59}
\eea

As in the case of the bosonic string, the trace in (\ref{strace})
can be represented by an integral over the transverse coordinate of the 
string, when the world sheet is a torus. 
The Green-Schwartz action on a toroidal world-sheet is~ \cite{gsw}
\be
S_{l.c.}=-\frac{1}{4\pi\alpha'}\int\d^2\sigma
\sqrt{g} g^{\alpha\beta}\left(
\partial_\alpha \vec X\cdot \partial_\beta \vec X-i2\pi\alpha'\bar\psi^i
\gamma_\alpha\partial_\beta\psi^i\right)
\ee
where $\gamma_\alpha=e^a_\alpha \gamma_a$, with $e^a_\alpha$ the
$zweibein$, and $\gamma_a$ are two dimensional Dirac matrices.
The world-sheet spinors $\psi^i$ belong to the ${\bf 8}_s$ 
representation of  $SO(8)$ and $g_{\alpha\beta}$ is given in (\ref{metric}).
The superstring free energy density then reads
\begin{eqnarray}
\frac{F}{L}=-
\sum_{\stackrel {n=1} {n ~\rm odd}}^\infty 
\int^{1/2}_{-1/2}\frac{d\tau_1}{4\pi^2\alpha'}
\int_0^\infty \frac{d\tau_2}{\tau_2^2}~
e^{-\frac{\beta^2 n^2}{4\pi\alpha'\tau_2}}\qquad\qquad\qquad \cr  \cdot~
\int d\vec X d\psi^i \exp\left[-\frac{1}{4\pi\alpha'}
\int d^2\sigma\sqrt{g}
g^{\alpha\beta}\left(
\partial_\alpha \vec X\cdot \partial_\beta \vec X-i2\pi\alpha'\bar\psi^i
\gamma_\alpha\partial_\beta\psi^i \right)\right]
\end{eqnarray}

\subsubsection{Equivalence to NSR formulation}

The type II superstring free energy in the Neveu-Schwarz-Ramond formulation
of superstring theory was discussed, for example, by Atick and Witten 
\cite{aw} who obtained the free energy density 
\begin{eqnarray}
\frac{F}{V}&=&\frac{1}{4}\left(\frac{1}{4\pi^2\alpha'}\right)^5\int_{\cal F}
\frac{d^2\tau}{\tau^6_2}\left|\frac{1}{\eta(\tau)}\right|^{24}
\sum_{n,m=-\infty}^{+\infty}
e^{-\beta^2(n^2+m^2|\tau|^2-2\tau_1 nm)/(4\pi\alpha'\tau_2)}\cr
&&\left[\left(\theta_2^4\bar\theta_2^4+\theta_3^4\bar\theta_3^4+
\theta_4^4\bar\theta_4^4\right)(0,\tau)+e^{i\pi(n+m)}
\left(\theta_2^4\bar\theta_4^4+\theta_4^4\bar\theta_2^4\right)(0,\tau)\right.\cr
&& \left.-e^{i\pi n}
\left(\theta_2^4\bar\theta_3^4+\theta_3^4\bar\theta_2^4\right)(0,\tau)
-e^{i\pi m}
\left(\theta_3^4\bar\theta_4^4+\theta_4^4\bar\theta_3^4\right)(0,\tau)\right]
\end{eqnarray}
where the Teichm\"uller parameters are integrated over the region ${\cal F}$,
which is the fundamental domain of the torus.
In this expression the sum over $m$ and $n$ is the sum over maps of the torus
in which the $space$ and $time$ time coordinates, respectively, of the torus 
wrap the target space $S^1$ $m$ and $n$ times.
One can use modular transformations to characterise these wrappings by a single
integer.  In doing so, the integration domain for the Teichm\"uller parameters
is expanded from ${\cal F}$ to the region ${\cal S}$, $-1/2<\tau_1<1/2$,\  
$0<\tau_2<\infty$.

We set $m=0$ and integrate 
over the region ${\cal S}$ in the above expression to get
\begin{equation}
\frac{F}{V}=\frac{1}{4}\left(\frac{1}{4\pi^2\alpha'}\right)^5
\int^{1/2}_{-1/2} d\tau_1\int_0^\infty
\frac{d\tau_2}{\tau^6_2}\left|\frac{1}{\eta(\tau)}\right|^{24}
\sum_{n=-\infty}^{+\infty}
e^{-\beta^2 n^2/(4\pi\alpha'\tau_2)} C(n,\tau)
\label{Faw}
\end{equation}
where $C(n,\tau)$ is the following combination of thetas. 
\begin{eqnarray}
C(n,\tau)&=&\left[\left(\theta_2^4\bar\theta_2^4+\theta_3^4\bar\theta_3^4+
\theta_4^4\bar\theta_4^4-
\theta_3^4\bar\theta_4^4-\theta_4^4\bar\theta_3^4\right)(0,\tau)\right.\cr
&+& \left.e^{i\pi n}
\left(\theta_2^4\bar\theta_4^4+\theta_4^4\bar\theta_2^4
-\theta_2^4\bar\theta_3^4-\theta_3^4\bar\theta_2^4\right)(0,\tau)\right]
\end{eqnarray}
$C(n,\tau)$ can be rearranged according to
\begin{equation}
C(n,\tau)=\left(\theta_3^4(0,\tau)-\theta_4^4(0,\tau)-
e^{i\pi n}\theta_2^4(0,\tau)\right)\left(\bar\theta_3^4(0,\tau)-
\bar\theta_4^4(0,\tau)-e^{i\pi n}\bar\theta_2^4(0,\tau)\right) 
\end{equation}
Using Jacobi's identity (\ref{abstruse}) 
$C(n,\tau)$ becomes
\begin{equation}
C(n,\tau)=2\theta_2^4\bar\theta_2^4(0,\tau)(1-e^{i\pi n})=
2\left|\theta_2(0,\tau)\right|^8(1-e^{i\pi n})
\end{equation}
The free energy (\ref{Faw}) then reads
\begin{equation}
\frac{F}{V}=\left(\frac{1}{4\pi^2\alpha'}\right)^5
\int^{1/2}_{-1/2} d\tau_1\int_0^\infty
\frac{d\tau_2}{\tau^6_2}\left|\frac{1}{\eta(\tau)}\right|^{24}
2\sum_{\stackrel{n=1}{n\ odd}}^{+\infty}
e^{-\frac{\beta^2 n^2}{4\pi\alpha'\tau_2}}
\left|\theta_2(0,\tau)\right|^8
\label{Faw1}
\end{equation}

The product representation of $\theta_2(0,\tau)$ in (\ref{triple})
implies that
\begin{equation}
\left|\frac{1}{\eta(\tau)}\right|^{24} \left|\theta_2(0,\tau)\right|^8
=2^8\prod_{n=1}^{\infty}
\left|\frac{1+e^{2\pi i n\tau}}{1-e^{2\pi i n\tau}}\right|^{16}
\equiv 2^8 \left|\theta_4\left(0,2\tau\right)\right|^{-16}
 \end{equation}
and eq.(\ref{Faw1}) is identical with (\ref{59})).

\section{The thermodynamic partition function of matrix string theory}

\subsection{Matrix string Hamiltonian}

In this section we derive the matrix string Hamiltonian starting from the 
matrix theory formalism~\cite{bfss}. In particular we wish to identify 
the dependence on the string coupling constant $g_s$, on the string tension
$\alpha'$.
The starting point is the matrix theory Hamiltonian, which describes a stack
of $N$ D0-branes.   In terms of the ten dimensional superstring 
string tension~\cite{wati} it has the form
\be
H=\frac{1}{2 g\sqrt{\alpha'}}\tr\left(g^2\alpha'\Pi_a^2-
\frac{1}{(2\pi\alpha')^2}\left[X^a,X^b\right]^2 -
\frac{1}{2\pi\alpha'}\psi^T\gamma_a\left[X^a,\psi\right]\right)
\label{hamsu}
\ee
where $a=1,\dots,9$. Each of the adjoint scalar matrices $X^a$ is a 
hermitian $N\times N$ matrix, where $N$ is the number of $0$-branes.
The super-partners of the $X$ fields are the 16 component spinors 
$\psi$ which transform under the $SO(9)$ Clifford algebra given by the 
16$\times$16 matrices $\gamma^a$. Conventionally $M-$theory is related to
type-IIA string theory via the compactification of the 11-th direction,
which relates the coupling constant $g$ to the compactification radius of
the 11-th dimension $R_{11}$ through
$g\sqrt{\alpha'}=R_{11}$.
The Hamiltonian (\ref{hamsu}) can also be expressed in 
eleven dimensional Planck units, by replacing $g\sqrt{\alpha'}=R_{11}$
and $\alpha'=l^2_p g^{-2/3}$, with $l_p$ the 11-dimensional Plank 
length
\be
H=\frac{R_{11}}{2}\tr\left(\Pi_a^2- 
\frac{1}{4\pi^2 l_p^6}\left[X^a,X^b\right]^2 -
\frac{1}{2\pi l^3_p} \psi^T\gamma_a\left[X^a,\psi\right]\right)\ .
\ee
The mass-dimensions of $H$ and of the fields are 
$$
[H]=M\quad,\qquad [X^a]=M^{-1}\quad,\qquad [\psi]=M^0\ .
$$
We now compactify the 9th dimension on a circle of radius $R_9$. After the 
usual $T$-duality, we can identify $X^9$ with the covariant derivative 
$$
X^9\longrightarrow i R_9 D_{1}
=i R_9(\frac{\partial}{\partial\sigma_{1}}-iA_{1})
$$ 
where the coordinate $\sigma_1\in [0,1]$. The momentum conjugate to $X^9$ 
will be identified with the electric field $E$ via $E=R_9 \Pi_9$.
This procedure~\cite{wati} gives the 
Hamiltonian (where $i=1,\dots,8$ now labels the transverse coordinates)
\bea
H=\frac{R_{11}}{2}\int_0^1d\sigma_1\tr\left[\Pi_i^2+\frac{R_9^2}{4\pi^2
l_p^6}(D_{1} X^i)^2 -\frac{1}{4\pi^2 l_p^6}\left[X^i,X^j\right]^2
+\frac{E^2}{R_9^2} \right.\cr\left. -  
\frac{1}{2\pi l_p^3}\psi^T\gamma_i\left[X^i,\psi\right]-
\frac{i R_9}{2\pi l_p^3}\psi^T \gamma\cdot D_{1}\psi\right]
\eea

To arrive at the matrix string point of view~\cite{dvv}, we now interchange 
the 9-th and the 11-th direction by defining the string scale 
$\sqrt{\alpha'}$ and the string coupling constant $g_s$ in terms 
of $R_9$ and $l_p$
\be
R_9=g_s\sqrt{\alpha'}\quad,\qquad l_p=g_s^{1/3}\sqrt{\alpha'}\ .
\ee
Note that the mass-dimension of  $g_s$ is 0.
With this substitution we obtain the final result for 
the Hamiltonian of the matrix string model of ref. \cite{dvv} (which we shall
refer to as DVV),
\bea
H=\frac{R_{11}}{2}\int_0^1d\sigma_1\tr\left[\Pi_i^2+\frac{1}{4\pi^2
\alpha'^2}(D_{1} X^i)^2 -\frac{1}{4\pi^2 \alpha'^3 g_s^2}
\left[X^i,X^j\right]^2
+\frac{E^2}{g_s^2\alpha'} \right.\cr\left. -  
\frac{1}{2\pi \alpha'^{3/2} g_s}\psi^T\gamma_i\left[X^i,\psi\right]-
\frac{i }{2\pi\alpha'}\psi^T \gamma\cdot D_{1}\psi\right]
\label{HDVV}
\eea
Here, $R_{11}$ is the radius of the dimension that must be compactified 
in order to obtain the matrix description of M-theory - in its original 
form the matrix model describes M-theory in a reference frame which has 
infinite momentum in the 11'th direction.   In the more sophisticated 
proposal of ref. \cite{su},  this compactified direction is the light-cone 
direction $R^+$.  It was shown by Seiberg \cite{ns} that, 
with certain assumptions, 
a boost to the frame with infinite momentum of a theory compactified on
a small circle $R_{11}$ is equivalent to one with the light cone compactified.
For a more recent discussion of these issues see \cite{bilal1,bilal2,hyun}.
Here, anticipating that we are actually describing DLCQ M-theory, 
we shall make the replacement
$$
R_{11}~~\longrightarrow~~R^+
$$
The hypothesis is that this model describes DLCQ M-theory with one
dimension compactified (to describe the DLCQ IIA superstring).
The canonical momenta which appear in (\ref{HDVV}) are normalised so
that they have the conventional canonical commutators,
\begin{eqnarray}
\left[ X^i_{ab}(\sigma), \Pi^j_{cd}(\sigma')\right]&=&
i\delta_{ad}\delta_{bc}\delta^{ij}\delta(\sigma-\sigma')  \cr
\left[ A_{ab}(\sigma), E_{cd}(\sigma')\right]&=&
i\delta_{ad}\delta_{bc}\delta(\sigma-\sigma')  \cr
\left\{ \psi_{ab}(\sigma), \psi_{cd}(\sigma')\right\}&=&\frac{1}{2}
\delta_{ad}\delta_{bc}\delta(\sigma-\sigma')
\end{eqnarray}
Note that $g_s^2\alpha'$ plays the role of the inverse square of the
Yang-Mills coupling constant: $g_s^2\alpha'=g^{-2}_{\rm Y.M.}$

\subsection{Matrix string free energy}

Using the  Hamiltonian (\ref{HDVV}) we shall 
now construct the thermal partition function $Z$ of the 
matrix string theory in the limit $g_s\to 0$. 
The prescription discussed in the previous sections
for the treatment of the thermal ensemble for systems in the light-cone 
frame, and the fact that $p^-=N/R^+$ lead us to write $Z$ as
\be
Z=\tr e^{-\beta P^0}=\tr e^{-\beta\left(P^++P^-\right)/\sqrt{2}}~=~
\sum^\infty_{N=0}~e^{-N\beta/\sqrt{2}R^+}~\tr\left\{e^{-\beta H/\sqrt{2}}
\right\}
\label{Z}
\ee
where $P^+=H$ is the matrix string theory Hamiltonian in (\ref{HDVV}).
The trace of $\exp(-\beta H/\sqrt{2})$ is to be taken over gauge invariant
states of the two-dimensional super-Yang-Mills theory.  This trace
has the standard path integral expression \cite{gpy}
\begin{equation}
Z[\beta]=\sum_{N=0}^\infty 
\int [dA_\alpha][dX^i][d\psi^a]~\exp\left(-\beta N/\sqrt{2}R^+-S_E[A,X^i,
\psi]\right)
\label{pf33}
\end{equation}
where $S_E$ is the Euclidean action 
\begin{eqnarray}
S_E=\frac{1}{2}\int_0^1 d\sigma_1\int_0^1 d\sigma_2\tr\left\{ 
\frac{\beta R^+}{(2\pi\alpha')^2\sqrt{2}}D_1 X^i D_1 X^i
+\frac{\sqrt{2}}{\beta R^+}D_2 X^i D_2 X^i+ \right.\cr\left. +
\frac{g_s^2\alpha'\sqrt{2}}{\beta R^+}
F_{\mu\nu}^2-\frac{\beta R^+}{(2\pi\alpha')^2\sqrt{2}\alpha'g_s^2}
\sum_{i<j}\left[ X^i,X^j\right]^2
-i\frac{\beta R^+}{\sqrt{2}2\pi\alpha'}\psi^T\gamma D_1\psi- \right.\cr\left. 
-i\psi^T D_2\psi- \frac{\beta R^+}{2\pi\alpha'\sqrt{2\alpha'}g_s}
\psi^T\gamma^i\left[X^i,\psi\right]
\right\}
\end{eqnarray}
and the covariant derivative is $D_\mu=\partial_\mu-i[A_\mu,...]$.  Here,
we have rescaled the time $\sigma_2$ so that the integration is over
a box of area one, $0\leq\sigma_\mu<1$, and $\beta$ appears
as a factor in various coupling constants.  Boundary conditions in the path
integral are
\begin{eqnarray}
A_\mu(\sigma_1+1,\sigma_2)&=&A_\mu(\sigma_1,\sigma_2) \cr
X^i(\sigma_1+1,\sigma_2)&=&X^i(\sigma_1,\sigma_2) \cr
\psi(\sigma_1+1,\sigma_2)&=&\psi(\sigma_1,\sigma_2) \cr
A_\mu(\sigma_1,\sigma_2+1)&=&A_\mu(\sigma_1,\sigma_2) \cr
X^i(\sigma_1,\sigma_2+1)&=&X^i(\sigma_1,\sigma_2) \cr
\psi(\sigma_1,\sigma_2+1)&=&-\psi(\sigma_1,\sigma_2) 
\label{matbc}
\end{eqnarray}
The anti-periodicity of the fermion field in the Euclidean time 
comes from taking the trace in (\ref{Z}).

The limit $g_s\rightarrow 0$ of this theory was formulated by DVV \cite{dvv}.
In this limit, the field configurations which have finite action are
related to diagonal field configurations by
\begin{eqnarray}
X^i(\sigma)&=&U(\sigma)X^i_D(\sigma)U^{-1}(\sigma) \cr
\psi(\sigma)&=&U(\sigma)\psi_D(\sigma)U^{-1}(\sigma) \cr
A_\mu(\sigma)&=&
iU(\sigma)\left(\partial_\mu-iA^D_\mu(\sigma)\right)U^{-1}(\sigma)
\label{diag}
\end{eqnarray}
where $X^i_D$, $\psi_D$ and $A^D_\mu$ are diagonal matrices.  The
fields $X^i$, $\psi$ and $A_\mu$ have the (anti-)periodic 
boundary conditions (\ref{matbc}). This implies that the diagonal
matrices have boundary conditions which are periodic up to 
permutations (and gauge transformations generated by the Cartan 
sub-algebra), 
\begin{eqnarray}
(A^D_\mu)_a(\sigma_1+1,\sigma_2)&=
&(A^D_\mu)_{P(a)}(\sigma_1,\sigma_2)+2\pi (n_\mu)_a +\partial_\mu\theta_a\cr
(X^i_D)_a(\sigma_1+1,\sigma_2)&=&(X^i_D)_{P(a)}(\sigma_1,\sigma_2) \cr
(\psi_D)_a(\sigma_1+1,\sigma_2)&=&(\psi_D)_{P(a)}(\sigma_1,\sigma_2) \cr
(A^D_\mu)_a(\sigma_1,\sigma_2+1)&=&(A^D_\mu)_{Q(a)}(\sigma_1,\sigma_2) 
+2\pi (m_\mu)_a+\partial_\mu\phi_a\cr
(X^i_D)_a(\sigma_1,\sigma_2+1)&=&(X^i_D)_{Q(a)}(\sigma_1,\sigma_2) \cr
(\psi_D)_a(\sigma_1,\sigma_2+1)&=&-(\psi_D)_{Q(a)}(\sigma_1,\sigma_2) 
\label{diagbc}
\end{eqnarray}
where $a=1,...,N$, $P(a)$ and $Q(a)$ are permutations and $n_\mu,m_\mu$ 
are integers.  Consistency
requires that the two permutations commute,
\begin{equation}
PQ=QP
\end{equation}
To compute the partition function, in the $g_s\rightarrow 0$ limit, 
we should now do the path integral 
(\ref{pf33}) over only the diagonal components of the matrix fields with
the action
\begin{eqnarray}
S_{\rm diag}=
\frac{1}{2}\int_0^1 d\sigma_1\int_0^1 d\sigma_2\sum_{a=1}^N
\left\{ 
\frac{\beta R^+}{(2\pi\alpha')^2\sqrt{2}}\partial_1 X^i_a \partial_1 X^i_a
+\frac{\sqrt{2}}{\beta R^+}
\partial_2 X^i_a \partial_2 X^i_a+ \right.\cr\left. +
\frac{g_s^2\alpha'\sqrt{2}}{\beta R^+}
\left(\partial_1A_{2a}-\partial_2 A_{1a}\right)^2
-i\frac{\beta R^+}{\sqrt{2}2\pi\alpha'}\psi^T_a\gamma \partial_1\psi_a- 
i\psi^T_a \partial_2\psi_a
\right\}
\end{eqnarray}
and with the boundary conditions (\ref{diagbc}). We should then sum over
topologically distinct configurations, characterised by the permutations
$P$ and $Q$ and by the integers $m_\mu,n_\mu$. We shall not discuss the 
validity of the 
assumptions leading to  this starting point in the present  
Paper (see discussions in \cite{wyn1,wyn2}).  
 
The partition function that we must compute thus decomposes into topological
sectors as
\be 
Z=\sum^\infty_{N=0}\frac{1}{N!}
e^{-N\beta/\sqrt{2}R^+}\sum_{P,Q} Z(P,Q)\ ,
\label{Zpq}
\ee  
(where we have suppressed the dependence on the integers in the boundary 
conditions for gauge fields).
Here, for each $N$ we have divided by contribution by the volume of the
Weyl group, $N!$, which reflects the fact that the eigenvalues are defined 
up to a global permutation. (From the M-theory point of view, this factor gives
Boltzmann statistics to the D0-branes.)
The possibility of introducing different weights in the sum over
pairs of commuting permutation has been discussed in~\cite{dadda}.
We shall show in what follows that the correct partition function for the type 
IIA string emerges from (\ref{Zpq}) only when all of the pairs $(P,Q)$
have the same weight.

\subsubsection{Gauge fields are irrelevant}

We first observe that, in all cases, when $g_s\rightarrow 0$ the gauge fields
are irrelevant.  Consider for the moment the QED partition function
\be
Z_{U(1)}=\int[dA_1][dA_2]\exp\left[-
\int_0^1 d\sigma_1\int_0^1 d\sigma_2\left(
\frac{g_s^2\alpha'}{\sqrt{2}\beta R^+} 
(\partial_1A_2-\partial_2 A_1)^2\right)\right]
\ee
where gauge fields are periodic up to integers and gauge transforms,
\bea
A_\mu(\sigma_1,\sigma_2+1)=A_\mu(\sigma_1,\sigma_2) + 2\pi m_\mu
+\partial_\mu \theta_a \ ,\cr
A_\mu(\sigma_1+1,\sigma_2)=A_\mu(\sigma_1,\sigma_2) + 2\pi n_\mu
+\partial_\mu \phi_a\ .
\label{lgt}
\eea
The N'th power of this partition function 
would arise as the gauge field contribution to the full partition function
in the topological sector where $P$ and $Q$ are the trivial
permutation.  (For a non-trivial permutation, though the details would be
slightly different, the arguments
below would still hold.)

We can un-twist this boundary conditions with a non-periodic 
gauge transformation to present the gauge field in terms of a 
periodic gauge field $\hat A_\mu$ according to
\bea
A_\mu(\sigma_1,\sigma_2)&=&\hat A_\mu(\sigma_1,\sigma_2)+2\pi n\sigma_1
\delta_\mu^2
\eea
The partition function becomes 
\bea
&&Z_{U(1)}
=\sum_{n=-\infty}^{\infty}\exp\left[-
\frac{g_s^2\alpha'n^2}{\sqrt{2}\beta R^+}\right]
\cdot\cr&\cdot&
\int[d\hat A_\mu]\exp\left[-
\int_0^1 d\sigma_1\int_0^1 d\sigma_2
\left(\frac{g_s^2\alpha'}{\sqrt{2}\beta R^+} 
(\partial_1\hat A_2-\partial_2 \hat A_1)^2\right)\right]
\eea
Fixing the gauge $\partial_2 \hat A_2=0$ and taking into account the 
Faddev-Popov determinant, the integration on the gauge fields can 
be performed and gives
\be
Z_{U(1)}=\left(\frac{g_s^2\alpha'}{\sqrt{2}\pi\beta R^+}\right)^{\frac{1}{2}}
\sum_{n_2=-\infty}^{\infty}\exp\left[
-\frac{g_s^2\alpha'(n_2)^2}{\sqrt{2}\beta R^+}\right]
\ee
Poisson re-summing
\be
Z_{U(1)}=\sum_{n_2=-\infty}^{\infty}
\exp\left[-\frac{\sqrt{2}\pi^2\beta R^+ (n^2)^2} {g_s^2\alpha'}\right]
\ee
In the $g_s\to 0$ limit only the $n_2=0$ sector survives and, 
for consistency with the approximations 
made above, the contribution of the Yang Mills term 
to the thermal partition function of the matrix string 
theory, should just be set to 1.  

\subsubsection{Generalities about permutations}

A discussion on the permutations of $N$ elements commuting
with a given permutation $P$ has been given in~\cite{dadda,kostov}.
It is useful at this point to review some of the salient points, for which
we follow the appendix of \cite{dadda}.  

Consider a fixed permutation $P$ of a set of $N$ elements.  
It can be decomposed into cycles which are subsets of the $N$ elements such
that the permutation interchanges elements cyclically 
inside each subset. For example, for $N=9$, the permutation which can be
denoted by
$$
\alpha=\left(\matrix{1&2&3&4&5&6&7&8&9\cr 6&4&1&2&5&3&8&9&7\cr}\right)
$$
has the three cycles $\alpha(1)=6$, $\alpha(6)=3$, $\alpha(3)=1$
and $\alpha(7)=8$, $\alpha(8(=9$, $\alpha(9)=7$, the two-cycle $\alpha(2)=4$, 
$\alpha(4)=2$ and the one-cycle $\alpha(5)=5$.  Thus it
can be decomposed into the product of disjoint cycles
$$
\alpha=(163)(24)(5)(789)
$$

Consider such a decomposition of $P$.
Suppose
that the number of cycles of length $k$ 
is $r_k$, so that
$N=\sum_k kr_k$.  Clearly, the minimum value of $k$ is one and the maximum 
is $N$.  (In the above example, $r_1=1,r_2=1,r_3=2$.)

Denote the elements of the set $\{1,2,...,N\}$ by the three index notation
based on how they transform under $P$:
\begin{equation}
a^{k,\alpha}_n  ~~(k=1,...,N);(\alpha=1,...,r_k):(n=1,...,k)
\end{equation}
This is the nth element of the $\alpha$'th cycle of length $k$.  The
action of the permutation on this element is then
\begin{equation}
P\left(a^{k,\alpha}_n\right)=a^{k,\alpha}_{n+1~{\rm mod}~k}
\label{permact}
\end{equation}

Let $Q$ be a permutation which commutes with $P$.  Consider its action on
the cycle $$\left(a_1^{k,\alpha},...,a_k^{k,\alpha}\right)$$ of $P$. 
Eq. (\ref{permact}) implies that
\begin{equation}
Q\left(P\left(a^{k,\alpha}_n\right)\right)
=Q\left(a^{k,\alpha}_{n+1~{\rm mod}~k}\right)
\end{equation}
Since $PQ=QP$, this equation can be written as
\begin{equation}
P\left(Q\left(a^{k,\alpha}_n\right)\right)
=Q\left(a^{k,\alpha}_{n+1~{\rm mod}\ k}\right)
\end{equation}
This means that
$$\left(Q\left( a_1^{k,\alpha}\right), ,...,
Q\left(a_k^{k,\alpha}\right)\right)$$
is a cycle of length $k$ in $P$, i.e. that
$$\left(Q\left( a_1^{k,\alpha}\right), ,... ,
Q\left(a_k^{k,\alpha}\right)\right)=
\left(a_s^{k,\pi_k(\alpha)},...,
a_{s+k-1~{\rm mod}~k}^{k,\pi_k(\alpha)}\right)$$
Hence, there exists a permutation $\pi_k(\alpha)$ of the $r_k$ elements 
of the set of cycles of length $k$ which is induced by $Q$.  Clearly, $Q$
commutes with $P$.

Any such permutation of the cycles of length $k$ for each $k$ corresponds to 
an acceptable element of $Q$.  For each of these permutations, there is an
additional piece of information - the ordering of a cycle is cyclic, so the
permutation between two cycles of length $k$ can also involve an element of 
the cyclic group (or, alternatively, the first element of the cycle $\alpha$ 
can be mapped onto any element of the cycle $\pi_k(\alpha)$).
This implies that there exist $r_k$ integers, $s(k,\alpha)\in\{1,2,..,k\}$
(with $\alpha\in 1,2,...,r_k$) such that
\begin{equation}
Q\left( a^{k,\alpha}_n\right)=
a^{k,\pi_k(\alpha)}_{n+s(k,\alpha){~\rm mod}~k}
\label{defnq}
\end{equation}

Thus, a permutation $Q$ which commutes with $P$ is completely determined
by  assigning for each value of $k$,
\begin{itemize}
\item{}A permutation $\pi_k\in S_k$ ( where the permutation group acting on the $r_k$ cycles of order $k$ in $P$)
\item{}A set of $r_k$ integers $s(k,\alpha)\in\{1,2,...,k)$
\end{itemize}
The permutation $Q$ is then defined by (\ref{defnq}).

It follows that the number of permutations $Q$ which commute with a given
permutation $P$ is
\be
\prod_k r_k!k^{r_k}\ .
\ee

\subsubsection{How permutations determine the world-sheet metric}

Consider a set of diagonal components of fields 
obeying the boundary conditions
\bea
\lambda_a(\sigma_1+1,\sigma_2)=\lambda_{P(a)}(\sigma_1,\sigma_2)\ ,\cr
\lambda_a(\sigma_1,\sigma_2+1)=(-1)^f\lambda_{Q(a)}(\sigma_1,\sigma_2)\ .
\label{pq}
\eea
where $f=0$ for a Bose field and $f=1$ for a fermion.
Consider those  which occur in the cycles
of length $k$ of $P$
\bea
\lambda_{a^{k\alpha}_n}(\sigma_1+1,\sigma_2)=\lambda_{a^{k\alpha}_{n+1}}
(\sigma_1,\sigma_2)\ ,\cr
\lambda_{a^{k\alpha_n}}(\sigma_1,\sigma_2+1)=(-1)^{f}
\lambda_{a^{k\pi(\alpha)}_{n+s}}(\sigma_1,\sigma_2)\ .
\label{pq1}
\eea
For each fixed $\alpha$, we fuse these fields together into
a single function which has the property
\bea
\lambda_{a^{k\alpha}}(\sigma_1+k,\sigma_2)=\lambda_{a^{k\alpha}}
(\sigma_1,\sigma_2)\ ,\cr
\lambda_{a^{k\alpha}}(\sigma_1,\sigma_2+1)=(-1)^{f}
\lambda_{a^{k\pi(\alpha)}}(\sigma_1+s(k,\alpha),\sigma_2)\ .
\label{pq2}
\eea
Then, we consider a cycle of the permutation $Q$, which must be a subset
of the $r_k$ k-cycles of $P$.  Say this cycle is of length $r$ where 
$1\leq r\leq r_k$.  Then we fuse $r$ of the above fields together to 
get the single field which has (anti-)periodic boundary conditions
\bea
\lambda_{a^{k,r}}(\sigma_1+k,\sigma_2)=\lambda_{a^{k,r}}
(\sigma_1,\sigma_2)\ ,\cr
\lambda_{a^{k,r}}(\sigma_1,\sigma_2+r)=(-1)^{fr}
\lambda_{a^{k,r}}(\sigma_1+s,\sigma_2)\ .
\label{pq3}
\eea  
where $s=\sum_\alpha s(k,\alpha)~{\rm mod}~k$ is the accumulated shift
for the $r$ elements in the cycle of $Q$.

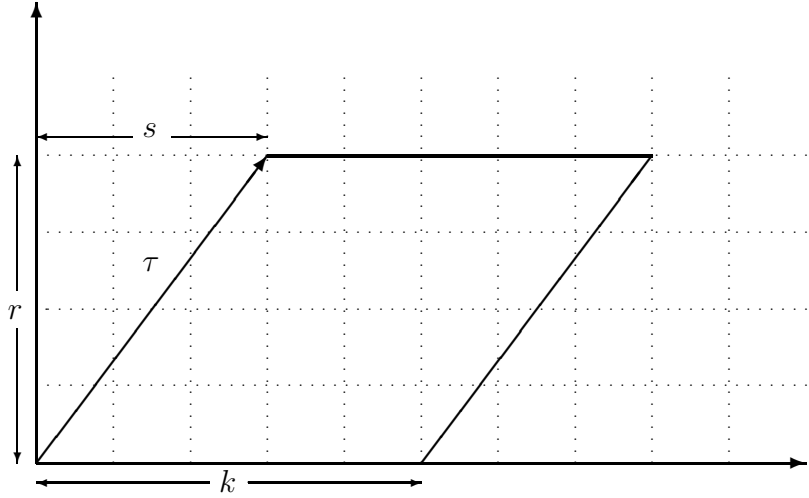
\begin{figure}[htb]
\begin{center}
\setlength{\unitlength}{0.00066700in}%
\begingroup\makeatletter\ifx\SetFigFont\undefined%
\gdef\SetFigFont#1#2#3#4#5{%
  \reset@font\fontsize{#1}{#2pt}%
  \fontfamily{#3}\fontseries{#4}\fontshape{#5}%
  \selectfont}%
\fi\endgroup%
\begin{picture}(6258,3858)(3376,-4186)
\thinlines
\put(3601,-2161){\makebox(1.8519,12.9630){\SetFigFont{5}{6}{\rmdefault}{\mddefault}{\updefault}.}}
\put(3601,-1636){\makebox(1.8519,12.9630){\SetFigFont{5}{6}{\rmdefault}{\mddefault}{\updefault}.}}
\put(4651,-1411){\vector( 1, 0){750}}
\put(4351,-1411){\vector(-1, 0){750}}
\put(5251,-4111){\makebox(1.8519,12.9630){\SetFigFont{5}{6}{\rmdefault}{\mddefault}{\updefault}.}}
\put(3451,-2611){\vector( 0, 1){1050}}
\put(3451,-2911){\vector( 0,-1){1050}}
\put(4951,-4111){\vector(-1, 0){1350}}
\put(5251,-4111){\vector( 1, 0){1350}}
\thicklines
\put(3601,-3961){\vector( 0, 1){3600}}
\put(3601,-3961){\vector( 1, 0){6000}}
\put(5401,-1561){\line( 1, 0){3000}}
\thinlines
\multiput(3601,-3361)(89.55224,0.00000){68}{\makebox(1.8519,12.9630){\SetFigFont{5}{6}{\rmdefault}{\mddefault}{\updefault}.}}
\multiput(4801,-961)(0.00000,-90.90909){34}{\makebox(1.8519,12.9630){\SetFigFont{5}{6}{\rmdefault}{\mddefault}{\updefault}.}}
\multiput(5401,-961)(0.00000,-90.90909){34}{\makebox(1.8519,12.9630){\SetFigFont{5}{6}{\rmdefault}{\mddefault}{\updefault}.}}
\put(3676,-2761){\makebox(1.8519,12.9630){\SetFigFont{5}{6}{\rmdefault}{\mddefault}{\updefault}.}}
\multiput(6001,-961)(0.00000,-90.90909){34}{\makebox(1.8519,12.9630){\SetFigFont{5}{6}{\rmdefault}{\mddefault}{\updefault}.}}
\put(4426,-2461){\makebox(0,0)[lb]{\smash{\SetFigFont{12}{14.4}{\rmdefault}{\mddefault}{\updefault}$\tau$}}}
\multiput(6601,-961)(0.00000,-90.90909){34}{\makebox(1.8519,12.9630){\SetFigFont{5}{6}{\rmdefault}{\mddefault}{\updefault}.}}
\multiput(7201,-961)(0.00000,-90.90909){34}{\makebox(1.8519,12.9630){\SetFigFont{5}{6}{\rmdefault}{\mddefault}{\updefault}.}}
\multiput(7801,-961)(0.00000,-90.90909){34}{\makebox(1.8519,12.9630){\SetFigFont{5}{6}{\rmdefault}{\mddefault}{\updefault}.}}
\multiput(8401,-961)(0.00000,-90.90909){34}{\makebox(1.8519,12.9630){\SetFigFont{5}{6}{\rmdefault}{\mddefault}{\updefault}.}}
\multiput(9001,-3961)(0.00000,90.90909){34}{\makebox(1.8519,12.9630){\SetFigFont{5}{6}{\rmdefault}{\mddefault}{\updefault}.}}
\multiput(3601,-2761)(89.55224,0.00000){68}{\makebox(1.8519,12.9630){\SetFigFont{5}{6}{\rmdefault}{\mddefault}{\updefault}.}}
\multiput(3601,-2161)(89.55224,0.00000){68}{\makebox(1.8519,12.9630){\SetFigFont{5}{6}{\rmdefault}{\mddefault}{\updefault}.}}
\multiput(3601,-1561)(89.55224,0.00000){68}{\makebox(1.8519,12.9630){\SetFigFont{5}{6}{\rmdefault}{\mddefault}{\updefault}.}}
\thicklines
\put(6601,-3961){\line( 3, 4){1800}}
\put(3601,-3961){\vector( 3, 4){1800}}
\thinlines
\multiput(4201,-961)(0.00000,-90.90909){34}{\makebox(1.8519,12.9630){\SetFigFont{5}{6}{\rmdefault}{\mddefault}{\updefault}.}}
\put(5026,-4186){\makebox(0,0)[lb]{\smash{\SetFigFont{12}{14.4}{\rmdefault}{\mddefault}{\updefault}$k$}}}
\put(3376,-2836){\makebox(0,0)[lb]{\smash{\SetFigFont{12}{14.4}{\rmdefault}{\mddefault}{\updefault}$r$}}}
\put(4426,-1411){\makebox(0,0)[lb]{\smash{\SetFigFont{12}{14.4}{\rmdefault}{\mddefault}{\updefault}$s$}}}
\end{picture}
\end{center}
\caption{Connected covering torus with $r=4$, $k=5$ and $s=3$.}
\label{fig1}
\end{figure}

The space on which coordinates which are arguments of the field in (\ref{pq3})
take values is the torus depicted in fig. 1.  The contribution to the path
integral of this set of fields is denoted by
\bea
Z(r,k,s)=
\int[dX^i][d\psi] \exp\left[-\frac{1}{2}
\int_0^r d\sigma_2\int_{\frac{s}{r}\sigma_2}^{k+\frac{s}{r}\sigma_2}
d\sigma_1\left( \frac{\beta R^+}{\sqrt{2}(2\pi\alpha')^2}
(\partial_1 X^i)^2\right.\right.
\cr\left.\left.+\frac{\sqrt{2}}{\beta R^+}(\partial_2 X^i)^2 -
i\frac{\beta R^+}{\sqrt{2}2\pi\alpha'}\psi^T
\gamma\cdot\partial_1\psi-i\psi^T\partial_2\psi
\right)\right]
\label{zrks}
\eea
where the integration region is the torus of fig.1 .
The boundary conditions for the Bose fields are
\begin{eqnarray}
X^i(\sigma_1+k,\sigma_2)=X^i(\sigma_1,\sigma_2)\cr
X^i(\sigma_1,\sigma_2+r)=X^i(\sigma_1+s,\sigma_2)
\end{eqnarray}
and for the Fermi fields are
\begin{eqnarray}
\psi(\sigma_1+k,\sigma_2)=\psi(\sigma_1,\sigma_2)\cr
\psi(\sigma_1,\sigma_2+r)=(-1)^r\psi(\sigma_1+s,\sigma_2)
\end{eqnarray}

It is possible to change the coordinates in the action (\ref{zrks}) so
that the integration region is the square torus $(\sigma_1, \sigma_2)\in
\left([0,1),[0,1)\right)$.  The appropriate coordinate transformation is
\bea
\sigma_1'&=&\frac{\sigma_1}{k} - \frac{s\sigma_2}{kr}\ ,\cr
\sigma_2'&=& \frac{\sigma_2}{r}
\eea
Then, the action becomes
\be
S=\frac{1}{4\pi\alpha'}
\int_0^1 d\sigma_2\int_{0}^{1}
d\sigma_1\sqrt{g}\left(
g^{\alpha\beta}
\partial_\alpha \vec X\cdot \partial_\beta \vec X-i 2\pi\alpha'\bar\psi^i
\gamma^a e^\alpha_a\partial_\alpha\psi^i \right)
\label{srks}
\ee
where 
\be
g_{\alpha\beta}=\left( \matrix{   1 & \tau_1\cr \tau_1& 
\vert\tau\vert^2\cr}\right)\ 
\ee
with
\be
\tau_1=\frac{s}{k}\quad,\quad\quad\quad\tau_2
=\frac{r\beta R^+}{2\sqrt{2}k\pi\alpha'}\quad .
\ee
and $e^\alpha_a$ a zweibein which corresponds to the metric $g_{\alpha\beta}$.
Now, the boundary conditions are
\begin{eqnarray}
X^i(\sigma_1+1,\sigma_2)=X^i(\sigma_1,\sigma_2)\cr
X^i(\sigma_1,\sigma_2+1)=X^i(\sigma_1,\sigma_2)\cr
\psi(\sigma_1+1,\sigma_2)=\psi(\sigma_1,\sigma_2)\cr
\psi(\sigma_1,\sigma_2+1)=(-1)^r\psi(\sigma_1,\sigma_2)
\end{eqnarray}
Note that the boundary condition for the Fermi field still depends on $r$.

When $r$ is {\bf even}, 
the fermion and boson have the same boundary conditions.  
These sectors are supersymmetric.  The mode expansion of both the fermions 
and bosons contain zero modes.  Functional integration over bosonic zero modes 
produces a factor of the infinite 
volume of $R^8$ and integration over the fermionic
zero mode produces a factor of zero.  If we were computing the Witten index, 
this product of infinity times zero, suitably regulated would yield the 
number of zero energy states of the supersymmetric theory.  However, here, 
we are computing an extensive thermodynamic variable - the free energy - 
from which we must extract a factor of the volume of the space in order to 
obtain the free energy
density.  In this case, sectors which contain fermion zero modes do not 
contribute.   

On the other hand, when $r$ is {\bf odd}, the fermions have 
anti-periodic boundary conditions - supersymmetry is broken by this 
boundary condition - and there are no fermion zero modes in the mode 
expansion. These sectors will survive and contribute to the partition 
function.  Thus,
\begin{equation}
Z(k,r,s)=0~{\rm when}~r~{\rm is ~ even}
\label{zermod}
\end{equation}

\subsubsection{How the first few orders work}

We shall now construct explicitly the first few terms 
contributing to the partition function (\ref{Zpq}).
The first term is obtained by considering the trivial permutations of 
the $N$ eigenvalues in both directions. 
The eigenvalues are then periodic in both directions and eq.(\ref{pq}) reads
\bea
\lambda_i(\sigma_1+1,\sigma_2)=\lambda_{i}(\sigma_1,\sigma_2)\ ,\cr
\lambda_i(\sigma_1,\sigma_2+1)=\lambda_{i}(\sigma_1,\sigma_2)\ .
\label{pqst}
\eea
The transverse partition function will then be given by the product of $N$ 
partition function defined on the square torus of side 1.
\be 
Z^{(1)}=\sum^\infty_{N=0}\frac{1}{N!}
e^{-\frac{N\beta}{\sqrt{2}R^+}}[Z(1,1,0)]^N.
\ee

The second order contribution to (\ref{Zpq}) is provided by summing
on all the permutation that permute any two eigenvalues,
and leave unchanged the other $N-2$.
We denote $\lambda_1(\sigma_1,\sigma_2)$ and 
$\lambda_2(\sigma_1,\sigma_2)$ the two permuted eigenvalues. There
are of course 4 pairs of permutations $(P,Q)$ of two eigenvalues 
and all commuting.
The contribution of the trivial permutations in both direction is 
already taken into account in $Z^{(1)}$, this would give in fact
a ``disconnected'' pair of permutations.

If $P=(1,2)$ and $Q=(1)(2)$ the boundary conditions (\ref{pq})
are
\bea
\lambda_i(\sigma_1+2,\sigma_2)=\lambda_{i}(\sigma_1,\sigma_2)\ ,\cr
\lambda_i(\sigma_1,\sigma_2+1)=\lambda_{i}(\sigma_1,\sigma_2)\ 
\eea
for $i=1,2$, and as in (\ref{pqst}) for the other $N-2$ eigenvalues.
Thus this pair of commuting permutation gives a contribution
$[Z(1,1,0)]^{N-2}\cdot Z(1,2,0)$.
Analogously $P=(1)(2)$ and $Q=(1,2)$ give the boundary conditions
\bea
\lambda_i(\sigma_1+1,\sigma_2)=\lambda_{i}(\sigma_1,\sigma_2)\ ,\cr
\lambda_i(\sigma_1,\sigma_2+2)=\lambda_{i}(\sigma_1,\sigma_2)\ 
\eea
for $i=1,2$
with a contribution to $Z_T$ given by $[Z(1,1,0)]^{N-2}\cdot Z(2,1,0)$.
For $P=(1,2)$ $Q=(1,2)$ we have
\bea
\lambda_i(\sigma_1+2,\sigma_2)=\lambda_{i}(\sigma_1,\sigma_2)\ ,\cr
\lambda_i(\sigma_1,\sigma_2+1)=\lambda_{i}(\sigma_1+1,\sigma_2)\ 
\eea
so that $r=1,k=2$ and $s=1$.
By considering that there are $\left(\matrix{N\cr 2}\right)$ pairs
and that is necessary to have at least $N=2$ to have a permutation of 
two eigenvalues, the second order contribution to the partition function
is thus given by
\be 
Z^{(2)}=\sum^\infty_{N=2}\frac{1}{N!}\left(\matrix{N\cr 2}\right)
e^{-\frac{N\beta}{\sqrt{2}R^+}}[Z(1,1,0)]^{N-2}
\left[Z(2,1,0)+Z(1,2,0)+Z(1,2,1)\right]
\ee
This term can be easily rearranged by shifting $N\to N+2$ as
\be 
Z^{(2)}=\sum^\infty_{N=0}\frac{1}{N!}
e^{-\frac{(N+2)\beta}{\sqrt{2}R^+}}
\frac{[Z(1,1,0)]^N}{2}\sum_{r|2}\sum_{s=0}^{2/r-1} Z(r,\frac{2}{r},s)\ ,
\ee
where $\sum_{r|N}$ means the sum over the divisors of N.
In an analogous way one can construct the third order contribution 
to the partition function $Z_T$, this can be obtained by considering 
the pairs of connected commuting permutations of three eigenvalues.
Taking the trivial permutation $P=(1)(2)(3)$ in the $\sigma_1$ direction,
the two permutations $Q=(1,2,3),(1,3,2)$ in the $\sigma_2$ direction 
each one forms a connected commuting pair with $P$. 
Any other permutation $Q$ commute with $P=(1)(2)(3)$
but the pair $(P,Q)$ would be in this case a disconnected pair whose
contribution to the partition function is already 
contained in the first two term of the expansion. It can be
easily seen that the values of $r,k$ and $s$ corresponding to the two pairs
$(P=(1)(2)(3),Q=(1,2,3),(1,3,2))$ are equal and given by $r=3,k=1,s=0$.
The other pairs of connected commuting permutations can be 
easily constructed to give Table 1.
\begin{table}[htbp]
\begin{center}
\caption{Pairs of commuting connected permutations of three eigenvalues.}
\label{3eigen}
\vspace{.1in}
\begin{tabular}{|cc|rrr|}\hline
$P$   &   $Q$ & $r$   &  $k$ & $s$ \\ \hline 
(1)(2)(3)  & (1,2,3)    & 3    &   1 & 0 \\
(1)(2)(3)  & (1,3,2)    & 3    &   1 & 0 \\
(1,2,3)    & (1)(2)(3)  & 1    &   3 & 0 \\
(1,3,2)    & (1)(2)(3)  & 1    &   3 & 0 \\
(1,2,3)    & (1,2,3)    & 1    &   3 & 1 \\
(1,3,2)    & (1,3,2)    & 1    &   3 & 1 \\
(1,2,3)    & (1,3,2)    & 1    &   3 & 2 \\
(1,3,2)    & (1,2,3)    & 1    &   3 & 2 \\ \hline
\end{tabular}
\end{center}
\end{table}

As can be seen from Table \ref{3eigen} there are 4 groups of 2  
pairs of $(P,Q)$ with the same values of $(r,k,s)$, to which 
correspond the same integration region in the action in (\ref{zrks})
or the same metric in (\ref{srks}).

There are $\left(\matrix{N\cr 3}\right)$ choices for three eigenvalues
and is necessary to have at least $N=3$ to have a permutation of 
three eigenvalues, the third order contribution to the partition function
then reads
\bea
Z^{(3)}&=&\sum^\infty_{N=3}\frac{1}{N!}\left(\matrix{N\cr 3}\right)
e^{-\frac{N\beta}{\sqrt{2}R^+}}
[Z(1,1,0)]^{N-3}\cdot\cr\cdot
&&2\left[Z(3,1,0)+Z(1,3,0)+Z(1,3,1)+Z(1,3,2)\right]\ .
\eea
By shifting $N\to N+3$ we get
\be 
Z^{(3)}=\sum^\infty_{N=0}\frac{1}{N!}
e^{-\frac{(N+3)\beta}{\sqrt{2}R^+}}
\frac{[Z(1,1,0)]^N}{3}\sum_{r|3}\sum_{s=0}^{3/r-1} Z(r,\frac{3}{r},s)\ .
\ee

Let us compute now the fourth order.
We should consider at this order both the pair
of commuting connected permutations of four eigenvalues and 
the permutations of two pairs of two eigenvalues.
Following the procedure illustrated above, for the 
commuting connected permutations of four eigenvalues
we can construct the Table \ref{4eigen}.

\begin{table}[htbp]
\begin{center}
\caption{Pairs of commuting connected permutations of three eigenvalues.}
\label{4eigen}
\vspace{.1in}
\begin{tabular}{|cc|rrr|}\hline
$P$   &   $Q$ & $r$   &  $k$ & $s$ \\ \hline 
 
(1)(2)(3)(4)  & (1,2,3,4)     & 4    &   1 & 0 \\
(1)(2)(3)(4)  & (1,2,4,3)     & 4    &   1 & 0 \\
(1)(2)(3)(4)  & (1,3,2,4)     & 4    &   1 & 0 \\
(1)(2)(3)(4)  & (1,3,4,2)     & 4    &   1 & 0 \\
(1)(2)(3)(4)  & (1,4,3,2)     & 4    &   1 & 0 \\
(1)(2)(3)(4)  & (1,4,2,3)     & 4    &   1 & 0 \\
(1)(2)(3)(4)  & (1,2,3,4)     & 4    &   1 & 0 \\
(1,2,3,4)     & (1)(2)(3)(4)  & 1    &   4 & 0 \\
(1,2,4,3)     & (1)(2)(3)(4)  & 1    &   4 & 0 \\
(1,3,2,4)     & (1)(2)(3)(4)  & 1    &   4 & 0 \\
(1,3,4,2)     & (1)(2)(3)(4)  & 1    &   4 & 0 \\
(1,4,3,2)     & (1)(2)(3)(4)  & 1    &   4 & 0 \\
(1,4,2,3)     & (1)(2)(3)(4)  & 1    &   4 & 0 \\
(1,2,3,4)     & (1,2,3,4)     & 1    &   4 & 1 \\
(1,2,4,3)     & (1,2,4,3)     & 1    &   4 & 1 \\
(1,3,2,4)     & (1,3,2,4)     & 1    &   4 & 1 \\
(1,3,4,2)     & (1,3,4,2)     & 1    &   4 & 1 \\
(1,4,3,2)     & (1,4,3,2)     & 1    &   4 & 1 \\
(1,4,2,3)     & (1,2,3,4)     & 1    &   4 & 1 \\
(1,2,3,4)     & (1,3)(2,4)    & 1    &   4 & 2 \\
(1,2,4,3)     & (1,4)(2,3)    & 1    &   4 & 2 \\
(1,3,2,4)     & (1,2)(3,4)    & 1    &   4 & 2 \\
(1,3,4,2)     & (1,4)(2,3)    & 1    &   4 & 2 \\
(1,4,3,2)     & (1,3)(2,4)    & 1    &   4 & 2 \\
(1,4,2,3)     & (1,2)(3,4)    & 1    &   4 & 2 \\
(1,2,3,4)     & (1,4,3,2)     & 1    &   4 & 3 \\
(1,2,4,3)     & (1,3,4,2)     & 1    &   4 & 3 \\
(1,3,2,4)     & (1,4,2,3)     & 1    &   4 & 3 \\
(1,3,4,2)     & (1,2,4,3)     & 1    &   4 & 3 \\
(1,4,3,2)     & (1,2,3,4)     & 1    &   4 & 3 \\
(1,4,2,3)     & (1,3,2,4)     & 1    &   4 & 3 \\
(1,2)(3,4)    & (1,3)(2,4)    & 2    &   2 & 0 \\
(1,2)(3,4)    & (1,4)(2,3)    & 2    &   2 & 0 \\
(1,3)(2,4)    & (1,2)(3,4)    & 2    &   2 & 0 \\
(1,3)(2,4)    & (1,4)(2,3)    & 2    &   2 & 0 \\
(1,4)(2,3)    & (1,2)(3,4)    & 2    &   2 & 0 \\
(1,4)(2,3)    & (1,3)(2,4)    & 2    &   2 & 0 \\
(1,2)(3,4)    & (1,3,2,4)     & 2    &   2 & 1 \\
(1,2)(3,4)    & (1,4,2,3)     & 2    &   2 & 1 \\
(1,3)(2,4)    & (1,2,3,4)     & 2    &   2 & 1 \\
(1,3)(2,4)    & (1,4,2,3)     & 2    &   2 & 1 \\
(1,4)(2,3)    & (1,2,3,4)     & 2    &   2 & 1 \\
(1,4)(2,3)    & (1,3,2,4)     & 2    &   2 & 1 \\ \hline
\end{tabular}
\end{center}
\end{table}
As can be seen from Table \ref{4eigen} there are 7 groups of 6  
pairs of $(P,Q)$ with the same values of $(r,k,s)$, to which 
correspond the same integration region in the action in (\ref{zrks})
or the same metric in (\ref{srks}).  
Taking into account that there are $\left(\matrix{N\cr 4}\right)$
groups of 4 eigenvalues and $\frac{N(N-1)(N-2)(N-3)}{8}$ pairs of 
pairs of eigenvalues, after the shift $N\to 4$ 
the fourth order contribution to the partition function reads
\bea
Z^{(4)}=\sum^\infty_{N=0}\frac{1}{N!}
e^{-\frac{(N+4)\beta}{\sqrt{2}R^+}}\frac{[Z(1,1,0)]^N}{4}~\cdot~
 \sum_{r|4}\sum_{s=0}^{4/r-1} Z(r,\frac{4}{r},s)
\cdot\cr\cdot\sum^\infty_{N=0}\frac{1}{N!}
e^{-\frac{(N+4)\beta}{\sqrt{2}R^+}}\frac{[Z(1,1,0)]^N}{2}
\left[\frac{1}{2}\sum_{r|2}\sum_{s=0}^{2/r-1} Z(r,\frac{2}{r},s)\right]^2\ .
\eea
Up to the fourth order $Z$ then reads
\bea
&&Z=\sum^\infty_{N=0}\frac{1}{N!}
e^{-\frac{N\beta}{\sqrt{2}R^+}}{[Z(1,1,0)]^N} \cdot\cr
&\cdot& \left\{1+\frac{e^{-\frac{\sqrt{2}\beta}{R^+}}}{2}
\sum_{r|2}\sum_{s=0}^{2/r-1} Z(r,\frac{2}{r},s)
+\frac{1}{2}
\left[\frac{e^{-\frac{\sqrt{2}\beta}{R^+}}}{2}
\sum_{r|2}\sum_{s=0}^{2/r-1} Z(r,\frac{2}{r},s)\right]^2\right.\cr
&+&\left.\frac{e^{-\frac{3\beta}{\sqrt{2}R^+}}}{3}
\sum_{r|3}\sum_{s=0}^{3/r-1} Z(r,\frac{3}{r},s)+
\frac{e^{-\frac{2\sqrt{2}\beta}{R^+}}}{4}
\sum_{r|4}\sum_{s=0}^{4/r-1} Z(r,\frac{4}{r},s)
\right\}
\eea
The pattern is quite clear, keeping into account also the higher orders,
each term in the square brackets containing 
sums over connected tori with the same area ($rk$),
is exponentiated with a weight given by
$$
\frac{e^{-\frac{r k\beta}{\sqrt{2}R^+}}}{rk}\ .
$$

\subsubsection{General solution}

It is clear from the discussion of the previous subsection that 
the free energy of the matrix string theory is given by the 
expression
\be
F=-\frac{1}{\beta}\log Z=-\frac{1}{\beta}
\sum_{N=0}^{\infty}\frac{e^{-\frac{N\beta}{\sqrt{2}R^+}}}{N}
\sum_{\stackrel{r|N}{r\ odd}}\sum_{s=0}^{N/r-1}Z
\left(r,\frac{N}{r},s\right)
\label{result}
\ee
where $Z(r,k,s)$ is given in eq.(\ref{zrks}). Note that $r$ gives the 
height of the torus, in the $\sigma_2$ direction and, taking into account 
eqn. ({\ref{zermod}) and the discussion before it, $r$ must be odd. 

With appropriate re-labeling of the integers, and evaluating the
functional integral (\ref{zrks}), this is identical to (\ref{lcpfs1}), the
expression for the free energy of the DLCQ type II superstring.
Note that, in comparing this with (\ref{lcpfs1}), rather than $1/k^2$ 
appearing in the summand in (\ref{result}), there is $1/N=1/kr$ where we 
label $k=N/r$.  This difference of a factor of $k/r$ is absorbed in a 
factor of $1/\tau_2$ in (\ref{lcpfs1}).

\section{Speculations}

We must caution the reader that one must be careful in 
interpreting thermodynamic quantities in a theory such as string theory
which intrinsically contains gravity.  Because of gravitational effects, 
one would not expect an exact, equilibrium thermal ensemble to exist.
However, to the extent that, at the basic level, 
both superstring theory and M-theory have some degree of space-time symmetry
such as time translation invariance, it is sensible to consider
the energy of quantum states and the thermodynamic partition function which
we consider simply as a spectral function which encodes the energies and
degeneracies of stationary quantum states.

One interesting feature of string theory is the existence of a limiting
temperature \cite{h}, which could be interpreted as evidence of a phase
transition \cite{aw,kog,sath}.  M-theory should also contain this feature.
Relevant to understanding it in M-theory is to consider the Hagedorn 
transition in the discrete light cone quantisation of the superstring, that is,
in the strongly coupled matrix model.  The relevant partition function
is (\ref{lcpfs1}) which re-display below:
$$
\frac{F}{V}=-\sum_{\stackrel {n=-\infty} {n ~\rm odd} }^\infty
\sum_{k=1}^\infty
\sum_{s=0}^{k-1}\frac{1}{k^2}\left( \frac{1}{4\pi^2\alpha'\tau_2}\right)^5
2^8 \left|\theta_4\left(0,\tau\right)\right|^{-16}
e^{-n^2\beta^2/4\pi\alpha'\tau_2}
$$
High temperature occurs for small values of $\beta$.  When $\beta$ is small 
compared to $\alpha'$, the region of summation with small values
of $\tau_2$ are important in the summand which behaves as
$$
\frac{1}{nk}
\exp(-n^2\beta^2/4\pi\alpha'\tau_2)\exp(2\pi/\tau_2)
~~~~~
$$
$$
~~~~~~~~~~~~~~~~\propto~
\frac{1}{nk}
\exp(-nk\beta\sqrt{2}/2 R^+)\exp(4\pi^2\alpha'\sqrt{2}k/n\beta R^+)
$$
The summation over $k$ diverges at  $T_H
=1/k_B\sqrt{8\pi^2\alpha'}$,
the Hagedorn temperature, independent
of the compactification radius $R^+$.  Since $k$ (see fig.1) is 
the spatial length of the string, this means that near the 
Hagedorn temperature, the partition sum is dominated by very long strings.

As was mentioned in the introductory section, the approach that we have 
followed in this Paper could be generalized to an expansion of the partition
function in powers of the string coupling constant, $g_s$.  What would have
to be shown is that the order $g_s^g$ correction to the free energy is
given by a summation over worldsheets with genus $g$.  This singular 
perturbation theory is similar to that encountered in the localization
formulae for path integrals describing integrable models \cite{local}.
Development of this perturbation theory
could well 
shed light on some aspects of the 
correspondence between gauge theory and $M$-theory supergravity 
contained in the Maldacena conjecture
(for recent reviews see~\cite{paolo,jens}).

\section*{Acknowledgements}
One of the authors (G.W.S.) acknowledges Andre Dubin for many 
informative conversations about the 
relationship between the permutation group and the representations
of Lie algebras.

\end{document}